\documentclass[journal]{IEEEtran}

\usepackage[utf8]{inputenc}
\usepackage[T1]{fontenc}

\usepackage[numbers,sort&compress]{natbib}

\usepackage{booktabs}
\usepackage{xcolor}
\usepackage{longtable}
\usepackage{graphics}
\usepackage{siunitx}
\usepackage{tabularx}
\usepackage{multirow}
\usepackage{amsmath}
\usepackage{xfrac}
\usepackage{dsfont}

\usepackage{graphicx}
\makeatletter
\let\MYcaption\@makecaption
\makeatother

\usepackage[font=footnotesize, labelformat=simple]{subcaption}

\makeatletter
\let\@makecaption\MYcaption
\makeatother

\usepackage[export]{adjustbox}

\usepackage[capitalise]{cleveref}

\usepackage[english]{babel}

\newcommand{\Var}[1]{\operatorname{Var}\left\{#1\right\}}
\newcommand{\E}[1]{\operatorname{E}\left\{#1\right\}}
\newcommand{\iu}{\mathrm{j}}

\title{A Nonlocal InSAR Filter for High-Resolution DEM Generation from TanDEM-X Interferograms}
\author{Gerald Baier, Cristian Rossi, Marie Lachaise, Xiao Xiang Zhu \IEEEmembership{Senior Member, IEEE}, Richard Bamler \IEEEmembership{Fellow, IEEE}
\thanks{This work is supported by the European Research Council (ERC) under the European Union's Horizon 2020 research and innovation programme (grant agreement no. ERC-2016-StG-714087, acronym: So2Sat, www.so2sat.eu) and the Helmholtz Association under the framework of the Young Investigators Group ``SiPEO" (VH-NG-1018, www.sipeo.bgu.tum.de).} \thanks{(\textit{Corresponding Author: Xiao Xiang Zhu})}
\thanks{G. Baier is with the Remote Sensing Technology Institute, German Aerospace Center, 82234 Wessling, Germany (e-mail: \mbox{gerald.baier@dlr.de}).}
\thanks{C. Rossi is with Satellite Applications Catapult, Harwell Campus, Didcot, OX11 0QR, United Kingdom.}
\thanks{M. Lachaise is with the Remote Sensing Technology Institute, German Aerospace Center, 82234 Wessling, Germany.}
\thanks{X. X. Zhu is with the Remote Sensing Technology Institute, German Aerospace Center, 82234 Wessling, Germany, and also with the Signal Processing for Earth Observation (SiPEO), Technical University of Munich, 80333 Munich, Germany. (e-mail: \mbox{xiao.zhu@dlr.de})}
\thanks{R. Bamler is with the Remote Sensing Technology Institute, German Aerospace Center, 82234 Wessling, Germany, and also with the Institute for Remote Sensing Technology, Technical University of Munich, 80333 Munich, Germany.}
}

\date{}

\begin{document}

\maketitle

\begin{abstract}

\textit{\textcolor{blue}{This is a preprint, to read the final version please go to IEEE Transactions on Geoscience and Remote Sensing on IEEE XPlore.}}

This paper presents a nonlocal InSAR filter with the goal of generating digital elevation models of higher resolution and accuracy from bistatic TanDEM-X strip map interferograms than with the processing chain used in production.
The currently employed boxcar multilooking filter naturally decreases the resolution and has inherent limitations on what level of noise reduction can be achieved.
The proposed filter is specifically designed to account for the inherent diversity of natural terrain by setting several filtering parameters adaptively.
In particular, it considers the local fringe frequency and scene heterogeneity, ensuring proper denoising of interferograms with considerable underlying topography as well as urban areas.
A comparison using synthetic and TanDEM-X bistatic strip map datasets with existing InSAR filters shows the effectiveness of the proposed techniques, most of which could readily be integrated into existing nonlocal filters.
The resulting digital elevation models outclass the ones produced with the existing global TanDEM-X DEM processing chain by effectively increasing the resolution from \SI{12}{\meter} to \SI{6}{\meter} and lowering the noise level by roughly a factor of two.

\end{abstract}

\begin{IEEEkeywords}
digital elevation model (DEM), interferometric synthetic aperture radar (InSAR), nonlocal filtering
\end{IEEEkeywords}

\section{Introduction}

With the global availability of the digital elevation model (DEM) produced by the German Aerospace Center's (DLR) TanDEM-X mission, topographic data with so far nonexistent spatial resolution and height accuracy have become accessible on a global scale.
The fact that a complete satellite mission was set in motion and executed for this sole purpose shows the demand and need for such data.
Even more attention should be paid not to compromise resolution and accuracy after acquisition by imperfect processing steps.

Phase denoising is a mandatory step within any InSAR DEM production workflow.
A more accurate phase estimate results not only in a less noisy DEM but also eases phase unwrapping.

Indiscriminate spatial averaging of the phase, also called boxcar multilooking, while being fast to compute and reducing the variance of the estimate, degrades resolution.
To address this issue, more advanced filtering methods have been the topic of research for more than two decades.
Lee's sigma filter and its later extensions~\cite{lee_sigma_filter_1983, lee_improved_sigma_filter_2009, lee_polsar_speckle_filtering_extended_sigma_2015} are examples of SAR and polarimetric SAR filters that include statistical tests for selecting pixels in the averaging process.

Nonlocal filters were first introduced for denoising optical images~\cite{buades_non-local_2005}.
In recent years, the have become increasingly popular within the denoising community, due to their unsurpassed noise reduction and detail preservation.
The foundation of their performance is a highly discriminate search for statistically homogeneous pixels, somewhat akin to the sigma filter, within a large area during the filtering process.
These features sparked research into adapting them to new domains, such as denoising regular SAR amplitude images~\cite{deledalle2009iwm, parrilli_nonlocal_2012, martino_scattering_based_nonlocal_2016}, interferograms~\cite{deledalle_nl-insar:_2011, lin_insar_tensor_svd_2015}, polarimetric SAR~\cite{chen_nonlocal_polsar_pretest_2011}, and a unified approach for SAR amplitude images, interferograms and polarimetric SAR images~\cite{deledalle_nl-sar:_2015}.
Recent publications applied the nonlocal filtering paradigm to SAR stacks in the fields of differential SAR interferometry~\cite{sica_nonlocal_multipass_2015} and 3D reconstruction using SAR tomography~\cite{hondt_nonlocal_tomosar_2018}.
The first nonlocal InSAR filter~\cite{deledalle_nl-insar:_2011} piqued our interest to produce DEMs from bistatic TanDEM-X strip map interferograms with improved resolution and accuracy compared to boxcar multilooking, which is employed in DLR's processing chain for the global TanDEM-X DEM\@.

For the original operational processing chain~\cite{breit_itp_2010, fritz_itp_2011, rossi_tandemx_rawdem_2012}, the need to cope with the data volume of the global DEM acquistion imposed severe design restrictions due to computational costs.
Boxcar multilooking was finally chosen, as the resulting DEM fulfills the TanDEM-X accuracy requirements~\cite{krieger_tandemx_2007} and its computational costs are negligible compared to the other processing steps.

Our research was motivated by the need for even higher-resolution DEMs which led DLR to commence research on the high-resolution DEM (HDEM) product, with increased horizontal resolution and vertical accuracy over selected areas~\cite{lachaise_eusar_tandemx_hdem_insar_update_2016} compared to the default TanDEM-X DEM product.
HDEMs rely on several new acquisitions with larger baselines resulting in smaller height errors from phase noise.
For comparison, the heights of ambiguity for HDEM range from \SI{10}{\meter} to \SI{20}{\meter} whereas the values for the regular DEM start at \SI{35}{\meter} and go up to \SI{50}{\meter}.
Thus, a boxcar averaging phase filter with a smaller spatial extent compared to the default processing toolchain suffices to fulfill the vertical accuracy goal and more of the original spatial resolution can be preserved.
\cref{tab:demspec} gives the specifications of the two available DEM products from DLR\@.

Our goal was to create a DEM similar in accuracy to the HDEM specifications by reprocessing the acquisitions made for the global TanDEM-X DEM\@.
The findings of our earlier investigation~\cite{zhu_improving_2014, zhu_nldem_2017} suggest that the qualities of nonlocal filters do indeed transfer to DEM generation.
We were able to produce a RawDEM, the initial DEM product used for creating the final TanDEM-X DEM, with \SI{6}{\meter} $\times$ \SI{6}{\meter} resolution showing more details and less noise compared to the operational product with a resolution of \SI{12}{\meter} $\times$ \SI{12}{\meter}.

Yet our straightforward application of NL-InSAR, the nonlocal filter introduced in~\cite{deledalle_nl-insar:_2011}, led to undesired terrace-like artifacts in the final DEM\@.
We also found, that the more recently published NL-SAR filter~\cite{deledalle_nl-sar:_2015} was unsuitable for DEM generation as it showed a tendency for oversmoothing.

\begin{table*}
\newcolumntype{Y}{>{\centering\arraybackslash}X}
\caption{Resolution and accuracy requirements of the standard global TanDEM-X DEM and the locally available HDEM~\cite{hoffmann_tdx_specs_2016}.}
\begin{tabularx}{\textwidth}{lYYY}
\toprule
& Independent pixel spacing
& Absolute horizontal and
& Relative vertical accuracy \\
&& vertical accuracies (\SI{90}{\percent})
& (\SI{90}{\percent} linear point-to-point) \\
\midrule
(global) TanDEM-X DEM
& \SI{12}{\meter} (\si{\ang{;;0.4}} at equator)
& \SI{10}{\meter}
& \SI{2}{\meter} (slope $\leq$ \SI{20}{\percent}) \\
&&&  \SI{4}{\meter} (slope $>$ \SI{20}{\percent}) \\
\midrule
(local) TanDEM-X HDEM
& \SI{6}{\meter} (\si{\ang{;;0.2}} at equator)
& \SI{10}{\meter}
& goal: \SI{0.8}{\meter} \\
&&& (\SI{90}{\percent} random height error)  \\
\bottomrule
\end{tabularx}%
\label{tab:demspec}
\end{table*}

This paper further elaborates on the issues we encountered when applying the nonlocal filtering paradigm to InSAR denoising and proposes a new nonlocal InSAR filter that takes these into consideration.
A key feature is its compensation of the deterministic, topographic phase component, which hampers the search for statistically homogeneous pixels in mountainous terrain.
It further factors in the diversity of natural terrain by using a local scene heterogeneity measure to select key filtering parameters instead of relying on a global, fixed set.
These techniques can readily be integrated into existing nonlocal InSAR filters to also bolster their performance.
A comparison with a LiDAR DEM gives an impression of and quantifies the level of improvement that can be achieved by employing nonlocal filters instead of conventional filters to real data.
Concerning the vastly increased computational cost, with the advances in semiconductor manufacturing processes and computing architecture, especially graphics processing units (GPUs), large-scale nonlocal filtering of SAR interferograms is nowadays feasible~\cite{baier_igarss_gpu_nonlocal_2016}.

The paper is structured as follows.
\cref{sec:nlinsar} briefly introduces the nonlocal filtering concept with respect to SAR interferometry.
The design decisions of the proposed filter are described in \cref{seg:nlswag} and are backed up by the experiments in \cref{sec:experiments}.
We discuss the impact of the new filter in \cref{sec:discussion} and conclude together with an outlook in \cref{sec:conclusion}.

\section{Nonlocal InSAR Filtering}%
\label{sec:nlinsar}

What sets nonlocal filters apart from other filters is the large area they operate over for denoising each pixel.
This area, called the search window, is inspected for similar pixels.
Their absolute position does not influence the later filtering process, true to their name “nonlocal”, unlike with many conventional neighborhood filters.
For detecting similar pixels, nonlocal filters do not only rely on comparing the pixel value alone, but also take their surrounding areas, henceforth referred to as patches, into account.
By doing so, textures, structures and features help identifying similar pixels and influence the filtering results to a far larger degree than with conventional filters.

\begin{figure}
    \centering
    \makeatletter%
    \if@twocolumn%
      \newcommand{\subfigwidth}{0.49\columnwidth}
    \else
      \newcommand{\subfigwidth}{0.3\columnwidth}
    \fi
    \begin{subfigure}[t]{\subfigwidth}
        \centering
        \includegraphics[width=\textwidth]{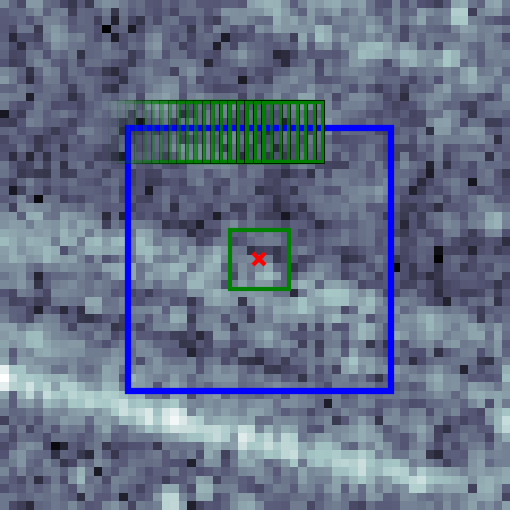}
        \caption{Search window (blue square) and patches (green squares)}%
        \label{subfig:nl_concept_data}
    \end{subfigure}
    \begin{subfigure}[t]{\subfigwidth}
        \centering
        \includegraphics[width=\textwidth]{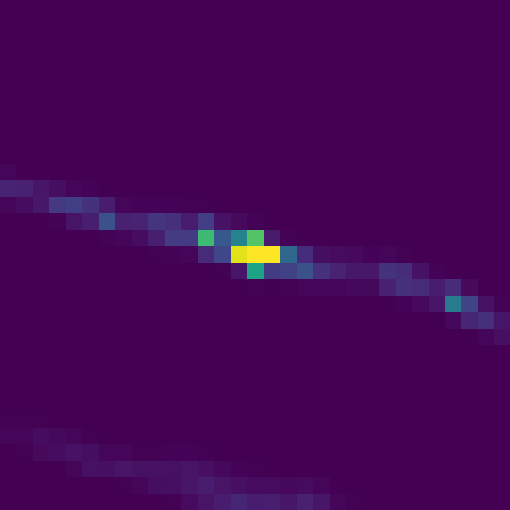}
        \caption{Similarity map}%
        \label{subfig:nl_concept_weight_map}
    \end{subfigure}
     \caption{The nonlocal filtering process: Inside the search window (blue square) centered at the pixel that is to be filtered \subref{subfig:nl_concept_data}, all pixels are checked for their similarity by comparing their surrounding patches to the center patch (green squares). The corresponding similarity map \subref{subfig:nl_concept_weight_map} shows that similar pixels are located along the edge.}%
     \label{fig:nl_concept}
\end{figure}

\cref{fig:nl_concept} illustrates this filtering process, where, in order to denoise the pixel marked by the red cross, all pixels inside the search window (blue square) are considered by comparing their surrounding patches to the center pixel's patch (all as green squares).
The resulting similarity map is depicted on the right and shows that the most similar pixels are located along the edge.

In the original version of the nonlocal filter, the Euclidean distance between patches was used as a measure of similarity.
This measure is the least square estimate for additive white Gaussian noise, a common and practical model for optical images.
As the noise characteristics of SAR profoundly differ, the earlier referenced filters for SAR, InSAR and polarimetric SAR all define similarity criteria depending on the statistics of the observed quantities: the speckle noise for SAR amplitude images, the interferometric phase for InSAR, or the covariance matrix for (Pol)(In)SAR\@.

The patch dissimilarities $\Delta$ in the search window are mapped into weights $w$ by a kernel.
In most cases, an exponential kernel or a slight adaption thereof is used

\begin{align}
    w = e^{-\frac{\Delta}{h}} \;,
    \label{eq:exp_kernel}
\end{align}
where $h$ sets the trade-off between filtering strength and detail preservation.
In the following, we assume that the weights are normalized to sum to one.
The estimate of an image $z$, in our case the interferogram

\begin{align}
    z = u_1 \bar u_2 = A_1 A_2 e^{\iu\varphi} = \lvert z \rvert e^{\iu\varphi}
\end{align}
of the master and slave images, at the pixel location $\mathbf{x}$ is computed as the weighted mean over the corresponding search window $\partial_\mathbf{x}$

\begin{align}
\hat z_\mathbf{x} = \sum\limits_{\mathbf{y} \in \partial_\mathbf{x}}  w_{\mathbf{x}, \mathbf{y}} z_\mathbf{y} \;.
\label{eq:weighted_mean}
\end{align}
The argument $\hat \varphi = \angle \hat z$ of $\hat z$ is the estimate of the true interferometric phase $\theta$.
In a similar fashion, estimates of the intensity

\begin{align}
    \hat I_\mathbf{x} &= \sum\limits_{\mathbf{y} \in \partial_\mathbf{x}}  w_{\mathbf{x}, \mathbf{y}} \frac{\lvert u_{1, \mathbf{y}} \rvert^2 + \lvert u_{2, \mathbf{y}}\rvert^2}{2}
    \label{eq:nl_int}
\end{align}
and coherence 

\begin{align}
    \hat \gamma_\mathbf{x} &= \frac{ \left\lvert \sum_{\mathbf{y} \in \partial_\mathbf{x}}  w_{\mathbf{x}, \mathbf{y}} u_{1, \mathbf{y}} \bar u_{2, \mathbf{y}} \right\rvert}{\sqrt{\sum_{\mathbf{y} \in \partial_\mathbf{x}}  w_{\mathbf{x}, \mathbf{y}} \lvert u_{1, \mathbf{y}} \rvert^2\sum_{\mathbf{y} \in \partial_\mathbf{x}}  w_{\mathbf{x}, \mathbf{y}} \lvert u_{2, \mathbf{y}} \rvert^2}}
    \label{eq:nl_coh}
\end{align}
can be obtained.
One can think of the nonlocal filter as a selector for statistically homogeneous pixels for the averaging process.

When dealing with SAR images and InSAR images in particular, there are several additional factors to consider when applying the nonlocal filter paradigm.
The next section highlights these pitfalls and describes how they are addressed specifically by the proposed method.

\section{Proposed Filter}%
\label{seg:nlswag}

\begin{figure}
    \centering
    \makeatletter%
    \if@twocolumn%
      \newcommand{\figwidth}{\columnwidth}
    \else
      \newcommand{\figwidth}{0.6\textwidth}
    \fi
    \makeatother
    \centering
    \includegraphics[width=\figwidth]{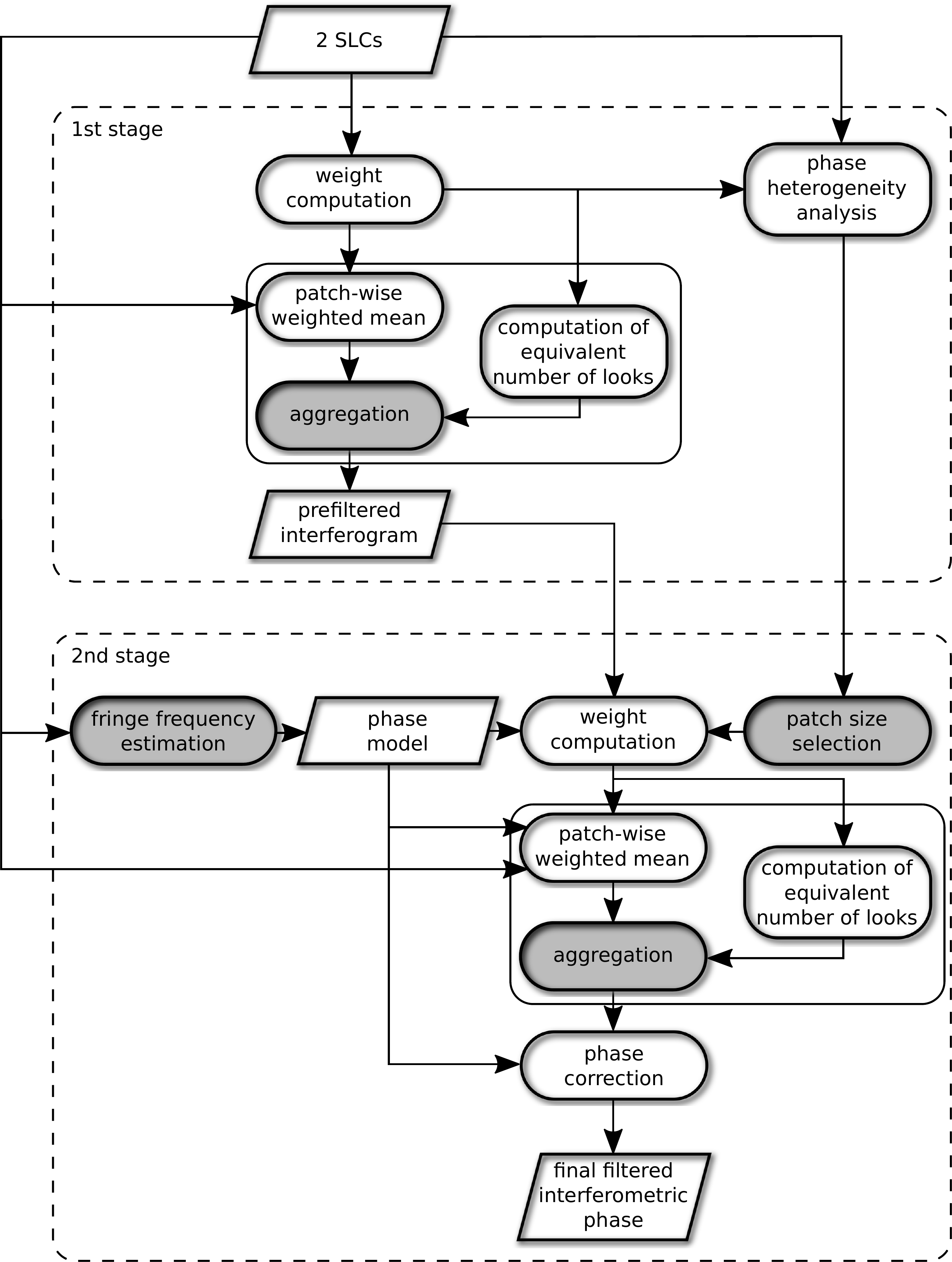}
    \caption{Flow graph of the proposed filter.
    Blocks that are highlighted in gray have their own respective subsections, which will also cover other related operations.
    The second stage uses the prefiltered output of the first stage for computing a new, more reliable set of weights.}%
    \label{fig:flowgraph}
\end{figure}

In the following, we will refer to the proposed method as NL-SWAG, short for \textbf{N}on\textbf{L}ocal-\textbf{S}AR interferogram filter for \textbf{w}ell-performing \textbf{A}ltitude Map \textbf{G}eneration.
\Cref{fig:flowgraph} shows a high-level flow graph of NL-SWAG\@.
The following paragraphs describe in greater detail the individual operations and how they affect the filtering performance and outcome.
We have highlighted in gray operations that are explicitly explained in the respectively named subsections, which will also cover other related blocks.

\subsection{Aggregation}

A common filtering artifact of nonlocal filters is the so-called \emph{rare patch effect}, which occurs when only few similar patches are located within the search window, resulting in subpar filtering performance.
The problem is especially prevalent near edges, as \cref{fig:nl_concept} illustrates, where for all patches that include the edge only few similar patches are found.
Aggregating multiple estimates is one approach to counter this behavior~\cite{lebrun_denoising_cuisine_2012}.

Instead of the traditional pixel-wise nonlocal means filter as in \cref{eq:weighted_mean}, NL-SWAG computes the patch-wise weighted mean

\begin{align}
    \mathbf{\hat z}_\mathbf{x} = \sum\limits_{\mathbf{y} \in \partial_\mathbf{x}}  w_{\mathbf{x}, \mathbf{y}} \mathbf{z}_\mathbf{y} \;.
\label{eq:weighted_mean_patch}
\end{align}
The overlapping patch estimates $\mathbf{\hat z}$ are then aggregated into a single pixel estimate, weighted by their equivalent number of looks $L$

\begin{align}
    \hat z_\mathbf{x} &= \frac{\sum_{\mathbf{y} \in \mathcal{P}_\mathbf{x}} L_\mathbf{y} \mathbf{\hat z}_\mathbf{y, x-y}}{\sum_{\mathbf{y} \in \mathcal{P}_\mathbf{x}} L_\mathbf{y}} \;,
\label{eq:aggregation}
\end{align}
where $\mathcal{P}_\mathbf{x}$ denotes the set of all pixel indices within a patch centered at $\mathbf{x}$ and $\mathbf{x-y}$ being the relative index inside the respective patch, i.e., $\mathbf{z}_\mathbf{y, x-y} = z_\mathbf{x}$.

The weighting by $L$ ensures that patch estimates with a higher number of looks, and therefore a smaller variance, have a larger impact on the final estimate.
The effective number of looks, i.e., the variance reduction of the weighted mean, can directly be computed from the weight map~\cite{deledalle_nl-insar:_2011}

\begin{align}
    L_\mathbf{x} = \frac{\left( \sum_{\mathbf{y} \in \partial_\mathbf{x}} w_{\mathbf{x, y}}\right)^2}{\sum_{\mathbf{y} \in \partial_\mathbf{x}} w^2_{\mathbf{x, y}}} \;.
\end{align}
Aggregation mitigates the rare patch effect as it also properly denoises pixels near features, such as edges, as long as they also belong to patches which do not contain said features.

\subsection{Two Stage Filtering}

SAR interferograms are affected by speckle and suffer from phase noise due to the innate coherence loss between two acquisitions, rendering the similarity estimates difficult and hereby degrading the denoising performance.

A solution that is often employed is a two-stage approach~\cite{dabov_bm3d_2007, deledalle_nl-sar:_2015, salmon_two-stage_denoising_2012}, where in the first step the so-called \emph{guidance image} is generated by prefiltering the input image.
In the second step, the guidance image is used to compute the patch similarities, which can now be more reliably estimated due to the reduced noise level.

The stages of NL-SWAG, which are also depicted in \cref{fig:flowgraph}, employ the two similarity criteria derived in~\cite{deledalle_nl-insar:_2011} for two single look complex images (SLC) in the first stage and a filtered interferogram in the second stage.

\subsubsection{First stage}

The similarity of two pixels in the first stage is the conditional likelihood of observing $u_{i,\mathbf{x}}$ and $u_{i, \mathbf{y}}$ ($i = 1,2$), given that the true parameters, the coherence $\gamma$, the intensity $I$ and the interferometric phase $\theta$ are identical~\cite{deledalle_nl-insar:_2011}:

\begin{multline}
p(u_{1, \mathbf{x}}, u_{1, \mathbf{y}}, u_{2, \mathbf{x}}, u_{2, \mathbf{y}} \vert I_\mathbf{x} = I_\mathbf{y}, \theta_\mathbf{x} = \theta_\mathbf{y}, \gamma_\mathbf{x} = \gamma_\mathbf{y}) =\\
\delta_{\mathbf{x}, \mathbf{y}}^1 = \sqrt{\frac{B}{C}}^3 \left( \frac{A+C}{A} \sqrt{\frac{C}{A-C}} - \arcsin \sqrt{\frac{C}{A}} \right),
\label{eq:pix_sim_likelihood}
\end{multline}
where

\begin{equation*}
\begin{aligned}
    A &= \left( A_{1,\mathbf{x}}^2 + A_{2,\mathbf{x}}^2+ A_{1,\mathbf{y}}^2 + A_{2,\mathbf{y}}^2  \right)^2 \;, \\
    B &= A_{1,\mathbf{x}} A_{2,\mathbf{x}} A_{1,\mathbf{y}} A_{2,\mathbf{y}} \;\textnormal{and} \\
    C &= 4 \left( A_{1,\mathbf{x}}^2 A_{2,\mathbf{x}}^2 + A_{1,\mathbf{y}}^2 A_{2,\mathbf{y}}^2 + 2 B \cos\left(\varphi_\mathbf{x} - \varphi_\mathbf{y} \right) \right) \;.
\end{aligned}
\end{equation*}
The patch similarity in the first stage is computed as

\begin{align}
    \Delta^1_{\mathbf{x}, \mathbf{y}} = \sum\limits_{\mathbf{o} \in \mathcal{O}} \log \delta^1_{\mathbf{x+o}, \mathbf{y+o}} \;,
    \label{eq:likelihood_patch_sim}
\end{align}
where $\mathcal{O}$ denotes the set of all index offsets in the patch.

The dissimilarities are mapped into weights by an exponential kernel as in \cref{eq:exp_kernel}.
As the purpose is only to reduce the noise level and remove outliers without introducing severe filtering artifacts before computing the similarities in the second step $h$ is set to a comparatively small value.
Except for the aggregation step, the first stage is identical to the non-iterative version of NL-InSAR and its guidelines for picking $h$ can be used.
The estimates of the phase, intensity and coherence are obtained via \cref{eq:weighted_mean}, \cref{eq:nl_int} and \cref{eq:nl_coh} together with the aggregation in \cref{eq:weighted_mean_patch} and \cref{eq:aggregation}.

\subsubsection{Second stage}
The second stage computes the similarities as a function of the coherence $\hat \gamma$, intensity $\hat I$ and interferometric phase $\hat \varphi$ estimates produced by the first stage.
The symmetric Kullback-Leibler divergence of two zero-mean complex circular Gaussian distributions, the underlying joint distribution of $\hat \gamma, \hat I$ and $\hat \varphi$, is given by~\cite{deledalle_nl-insar:_2011}

\makeatletter%
\if@twocolumn%
\begin{multline}
    \delta^2_{\mathbf{x}, \mathbf{y}} = \frac{4}{\pi} \left[ \frac{\hat I_\mathbf{x}}{\hat I_\mathbf{y}} \frac{1-\hat \gamma_\mathbf{x} \hat \gamma_\mathbf{y} \cos(\hat\varphi_\mathbf{x} - \hat\varphi_\mathbf{y})}{1-\hat \gamma_\mathbf{y}^2} \right. \\
                                      +               \left. \frac{\hat I_\mathbf{y}}{\hat I_\mathbf{x}} \frac{1-\hat \gamma_\mathbf{y} \hat \gamma_\mathbf{x} \cos(\hat\varphi_\mathbf{y} - \hat\varphi_\mathbf{x})}{1-\hat \gamma_\mathbf{x}^2} - 2 \right]
    \label{eq:kldivs}
\end{multline}
\else
\begin{equation}
    \begin{aligned}
        \delta^2_{\mathbf{x}, \mathbf{y}} = \frac{4}{\pi} \left[ \frac{\hat I_\mathbf{x}}{\hat I_\mathbf{y}} \frac{1-\hat \gamma_\mathbf{x} \hat \gamma_\mathbf{y} \cos(\hat\varphi_\mathbf{x} - \hat\varphi_\mathbf{y})}{1-\hat \gamma_\mathbf{y}^2} 
                                          +                      \frac{\hat I_\mathbf{y}}{\hat I_\mathbf{x}} \frac{1-\hat \gamma_\mathbf{y} \hat \gamma_\mathbf{x} \cos(\hat\varphi_\mathbf{y} - \hat\varphi_\mathbf{x})}{1-\hat \gamma_\mathbf{x}^2} - 2 \right] \;.
    \end{aligned}
    \label{eq:kldivs}
\end{equation}
\fi
\makeatother
and can be used as a similarity criterion.
Instead of a fixed patch size, the second stage changes the patch size adaptively based on the local heterogeneity.
The exact patch similarity and weight computation are covered in the following two sections since, as can be seen from \cref{fig:flowgraph}, it is based on other operations.

Even though the two-step approach alleviates the problems caused by the high noise level in SAR images, we have to stress that a repeated application of any filter can potentially introduce staircase-like artifacts in the filtered output as we observed with NL-InSAR\@.

To elaborate a little further: Just like traditional neighborhood filters, nonlocal filters can also be seen as diffusion filters~\cite{barash_framework_nonlinear_diffusion_bilateral_2004}.
Diffusion filters have the interesting property that their repeated application steadily decreases the noise level and produces piecewise constant approximations of the original data~\cite{weickert_review_nonlinear_diffusion_filtering_1997}.
While this can actually be a desired result for image segmentation or generating abstractions~\cite{winnemoeller_video_abstraction_2006}, for example, bilateral filters are often used to cartoonify photographs, in our case this phenomenon may lead to staircases in the generated DEM for iterative nonlocal algorithms, as errors of the phase estimate propagate and aggregate with every iteration.

\subsection{Patch Size Selection}

Patches contain information about the local texture and hence play a crucial role in distinguishing between suitable patches for averaging and patches that should be discarded.
That raises the question: How to select the best patch size?
In~\cite{duval_bias-variance_nl_means_2011}, the authors demonstrated that a global selection was suboptimal and that patch size should depend on the local neighborhood.
The following paragraphs repeat their reasoning and puts it into the context of SAR interferogram denoising for DEM generation.

For the original nonlocal filter, patch similarity, just like \cref{eq:likelihood_patch_sim}, is essentially the sum of all contained pixel similarities.
Naturally large patches reduce the variance and provide the most robust estimate of patch similarity.
This is indeed the best strategy for plains, agricultural fields or other slowly varying terrain.

The situation is quite different for more complex terrain, for instance urban sites or mountain ridges.
In these areas, a large patch size leads to the rare patch effect, since for every patch that contains some local structure only patches with similar features will have a significant impact on the averaging process.
The likelihood of finding such patches decreases with increasing patch size.

NL-SWAG's solution is to adaptively select the patch size as a function of local scene heterogeneity.
This way, a more robust patch similarity can be computed in flat regions or moderately hilly areas, due to the larger patch size, while at the same time the rare patch effect is alleviated in areas with many features and details.
Yet we have to stress that small patches come at the cost of less reliable patch similarity estimates.

We would further like to draw attention to the fact that the argument for an adaptive patch size selection to avoid the rare patch effect is identical to the one for aggregation.
Both measures favor patches that exclude local structures by either shrinking the patch or including estimates where the patch is moved off-center with respect to the pixel that is to be denoised.
This is somewhat contrary to the initial argument that patch-based methods perform so well because they take textures and details into account.
Patches indeed provide an effective mean for discarding patches of different classes.
But to maximize the number of patches that are classified as similar, both techniques also try to use the patch modification schemes we just mentioned.

To identify heterogeneous pixels and select the patch size accordingly we apply the local phase heterogeneity measure derived in~\cite{lee_insar_additive_1998}

\begin{align}
    \eta_\mathbf{x} = \frac{\Var{\varphi}_\mathbf{x} - \sigma^2_{0, \mathbf{x}}}{\Var{\varphi}_\mathbf{x}} \;,
  \label{eq:phase_linear_estimate}
\end{align}
which lies in the interval $\left[0, 1\right)$.
$\Var{\varphi}$ is the estimated variance of the phase in the search window and $\sigma^2_0$ the variance one would expected from the coherence~\cite{bamler_insar_1998}.
For non-heterogeneous terrain, $\Var{\varphi}$ is comparable in magnitude to $\sigma^2_0$ as only phase noise causes phase changes and \cref{eq:phase_linear_estimate} is close to $0$.
The situation changes when the search window contains structures, such as buildings.
Their distinct phase profiles increase $\Var{\varphi}$ resulting in larger phase heterogeneity values.

As the phase is wrapped, the filter first performs local unwrapping as in~\cite{lee_insar_additive_1998} to obtain the locally unwrapped phase $\tilde \varphi$ with respect to the average of the $5 \times 5$ pixels in the center.
The phase variance is then estimated inside the search window, weighted by the respective weight map computed in the first stage

\begin{align}
    \Var{\varphi}_\mathbf{x} &= \E{\varphi_\mathbf{x}^2} - \E{\varphi_\mathbf{x}}^2 \nonumber \\
                             &\approx \sum\limits_{\mathbf{y} \in \partial_\mathbf{x}} w_{\mathbf{x}, \mathbf{y}} \tilde \varphi_\mathbf{y}^2 - \left( \sum\limits_{\mathbf{y} \in \partial_\mathbf{x}} w_{\mathbf{x}, \mathbf{y}} \tilde \varphi_\mathbf{y} \right)^2 \;.
    \label{eq:phase_variance}
\end{align}
As $\Var{\varphi}$ is estimated in a local window due to insufficient sample size, \cref{eq:phase_linear_estimate} might be negative.
In this case, the heterogeneity measure is set to zero.

To yield a more reliable estimate of $\sigma_0$ the coherence is estimated following the methodology in~\cite{guarnieri_quick_dirty_coherence_estimator_1997} as

\begin{align}
    \gamma = \frac{\E{\lvert u_1 \rvert^2 \cdot \lvert u_2 \rvert^2}}{\sqrt{\E{\lvert u_1 \rvert^4} \E{\lvert u_2 \rvert^4}}} \;.
\end{align}
This way, the coherence is estimated from the speckle pattern and is not influenced by the topographic phase, which would yield an underestimation of the coherence if the common coherence estimator is used.
Just like in \cref{eq:phase_variance}, the expected value is replaced by the weighted mean over the respective quantities.

\begin{figure}
    \centering
    \makeatletter%
    \if@twocolumn%
      \newcommand{\subfigwidth}{0.49\columnwidth}
    \else
      \newcommand{\subfigwidth}{0.32\textwidth}
    \fi
    \makeatother
    \begin{subfigure}[t]{\subfigwidth}
        \includegraphics[width=\textwidth]{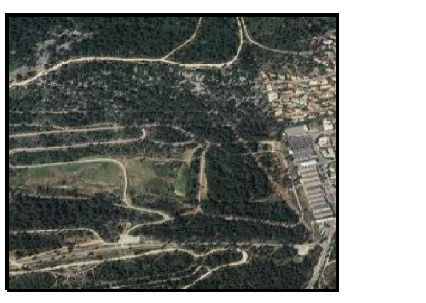}
        \caption{Optical image \textcopyright~Google~~~~~~~}%
        \label{subfig:phase_lhi_optical}
    \end{subfigure}
    \begin{subfigure}[t]{\subfigwidth}
        \includegraphics[width=\textwidth]{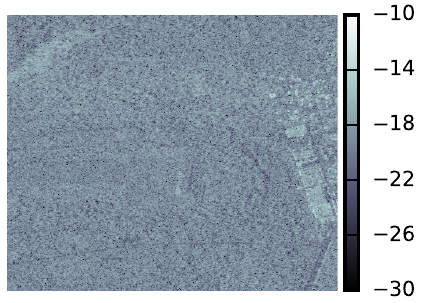}
        \caption{Master amplitude in \si{\decibel}}%
        \label{subfig:phase_lhi_ints}
    \end{subfigure}

    \begin{subfigure}[t]{\subfigwidth}
        \includegraphics[width=\textwidth]{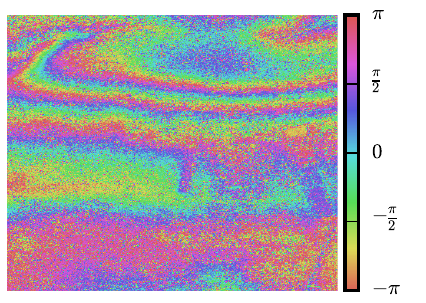}
        \caption{Phase}%
        \label{subfig:phase_lhi_phase}
    \end{subfigure}
    \begin{subfigure}[t]{\subfigwidth}
        \includegraphics[width=\textwidth]{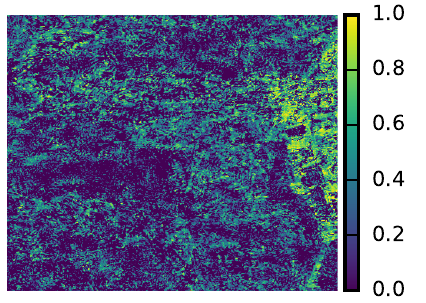}
        \caption{Phase heterogeneity}%
        \label{subfig:phase_lhi}
    \end{subfigure}
    \caption{Phase heterogeneity computed in the first stage. Urban areas, forests and grassland show different levels of heterogeneity.}%
    \label{fig:phase_lhi}
\end{figure}

An example of the heterogeneity measure is depicted in \cref{fig:phase_lhi}.
The urban area is clearly detected as being heterogeneous, the grassland is classified as the most homogeneous site and the forested areas are identified as moderately heterogeneous regions.

Instead of selecting a fixed patch size from a predefined set, depending on the local heterogeneity, NL-SWAG employs Gaussian windows of variable width.
A possible mapping of the phase heterogeneity index into Gaussian window widths could be

\begin{align}
    \sigma_{\textnormal{Gauss}} = 2\cdot(1-\eta) + 1 \;,
    \label{eq:lhi2gauss}
\end{align}
which gives strict lower and upper bounds for the window widths and is used in the remaining of the paper.
Other mappings would also be possible as long as they result in wide Gaussian windows for homogeneous areas and the reverse for heterogeneous areas.

As an alternative approach for selecting the best effective patch size, the phase variance in \cref{eq:phase_linear_estimate} could be computed in Gaussian windows of successively increasing widths.
This process is halted as soon as the heterogeneity level exceeds a predefined threshold, i.e., when significant phase changes, which most likely are the result of heterogeneous structures inside the patch, are detected.
A similar approach was presented in\ \cite{kervrann_adaptive_search_window_2008} for adaptively selecting the search window size.

For Gaussian blurring, the reduction in variance is related to $\sigma_\textnormal{Gauss}$ by approximately $4 \pi \sigma_\textnormal{Gauss}^2$.
So with \cref{eq:lhi2gauss}, the variance of the patch similarity estimation is reduced by a factor ranging from $4 \pi $ to $36 \pi$, roughly equivalent to $3 \times 3$ up to $11 \times 11$ patches.

Correspondingly to \cref{eq:likelihood_patch_sim}, the adaptive patch similarities are computed as the sum over the pixel similarities weighted by a Gaussian window $g_\mathbf{x}$

\begin{align}
    \Delta^2_{\mathbf{x}, \mathbf{y}} = \frac{\sum_{\mathbf{o} \in \mathcal{O}} g_{\mathbf{x}, \mathbf{o}} \delta^2_{\mathbf{x+o}, \mathbf{y+o}}}{\sum_{\mathbf{o} \in \mathcal{O}} g_{\mathbf{x}, \mathbf{o}}}  \;.
\end{align}

The patch dissimilarities still need to be mapped into weights, which in the second stage is also done by an exponential kernel.
We now face the problem how to select the normalization factor $h$ to compromise between bias and variance reduction.

\begin{figure}
\centering
\includegraphics{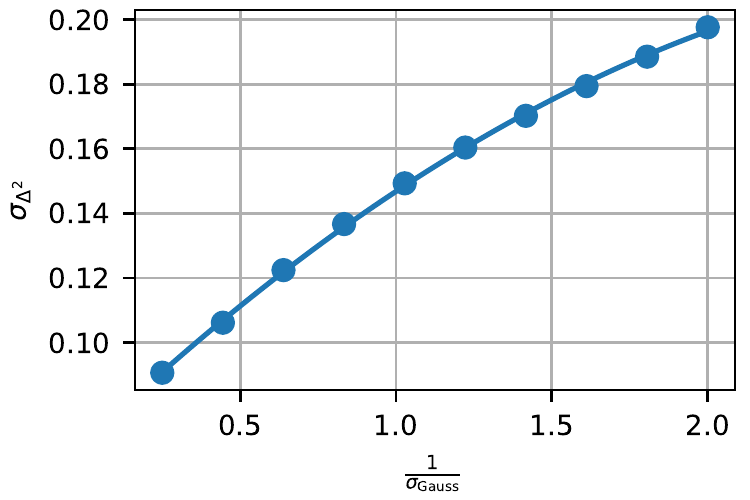}
\caption{Relationship between the width of the Gaussian window $\sigma_\textnormal{Gauss}$ and the standard deviation of the resulting patch similarities $\sigma_{\Delta^2}$. Due to the correlation of the pixel similarities there is no linear mapping.}%
\label{fig:polyfit}
\end{figure}

The standard deviation of $\Delta^2$ is reciprocally proportional to $\sigma_\textnormal{Gauss}$, which effectively governs the patch size.
Consequently, a fixed $h$ for all heterogeneity levels will be insufficient and a method is needed that accounts for varying patch sizes.
For this purpose, we selected a homogeneous training area and analyzed how the patch similarity's standard deviation $\sigma_{\Delta^2}$ changed with varying $\tfrac{1}{\sigma_\textnormal{Gauss}}$.
\cref{fig:polyfit} shows the relationship for a fixed set of Gaussian window widths at a homogeneous test site without any topography.
Clearly, the relationship is non-linear, due to the correlation between pixel similarities, but a second order polynomial, also depicted, is a good fit.

The weights are computed as
\begin{align}
    w_\mathbf{x, y} = \exp \left\{ -\frac{\Delta^2_{\mathbf{x}, \mathbf{y}}}{h \cdot \xi\left(\sigma^{-1}_{\textnormal{Gauss}, \mathbf{x}}\right)} \right\} \;,
\end{align}
where $\xi$ is the second order polynomial that accounts for the varying effective patch sizes and $h$ provides a fixed compromise between detail preservation and noise reduction.
In our experiments, we found that the interval $\left[ 1 \leq h \leq 2 \right]$ provided the best trade-off.

To account for the fact that, due to the Gaussian window, not every pixel in the patch estimate contributed equally to the similarity computation in contrast to \cref{eq:aggregation} the respective pixels are additionally weighted by their Gaussian weight in the final aggregation step

\begin{align}
    \hat z_\mathbf{x} &= \frac{\sum_{\mathbf{y} \in \mathcal{P}_\mathbf{x}} L_\mathbf{y}  g_{\mathbf{y}, \mathbf{x-y}} \mathbf{\hat z}_\mathbf{y, x-y}}{\sum_{\mathbf{y} \in \mathcal{P}_\mathbf{x}} L_\mathbf{y}  g_{\mathbf{y}, \mathbf{x-y}}} \;.
\label{eq:aggregation_fringe}
\end{align}

\subsection{Fringe frequency estimation and compensation}

\begin{figure}
    \centering
    \makeatletter%
    \if@twocolumn%
      \newcommand{\figwidth}{0.8\columnwidth}
    \else
      \newcommand{\figwidth}{0.5\textwidth}
    \fi
    \makeatother
    \includegraphics[width=\figwidth]{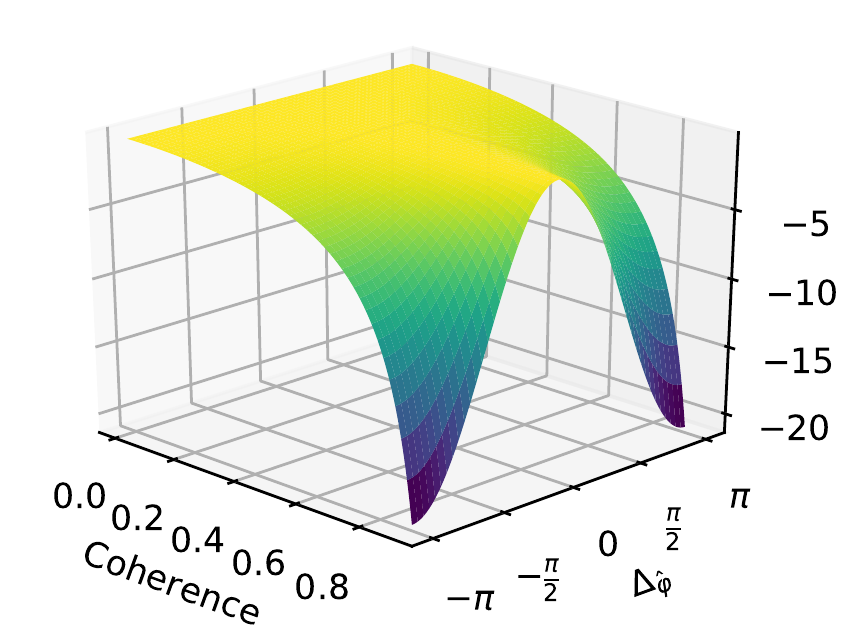}
    \caption{Symmetric Kullback-Leibler divergence from \Cref{eq:kldivs} for two pixels with identical reflectivity and coherence, dependent on their phase difference.}%
    \label{fig:kldivs}
\end{figure}

Another obstacle hindering the use of nonlocal InSAR filters for DEM generation is the actual topography, which, together with the atmosphere, the deformation and noise, contributes to the measured interferometric phase.
For the bistatic case, the acquisition mode of TanDEM-X interferograms for the generation of the global DEM, the deformation and the atmosphere components can be ignored, so that only the topography and noise components affect the similarity measure.
Due to the topographic phase component it is considerably harder to detect statistically homogeneous pixels in regions with non-negligible height differences, that is pixels with identical noise distribution but different heights.
\cref{fig:kldivs} shows the symmetric Kullback-Leibler divergence from \cref{eq:kldivs} with $\hat I_\mathbf{x} = \hat I_\mathbf{y}$ and $\hat \gamma_\mathbf{x} = \hat \gamma_\mathbf{y}$ as a function of the coherence and the phase difference $\Delta_{\hat \varphi} = \hat \varphi_\mathbf{x} - \hat \varphi_\mathbf{y}$, that is used in the second stage as the similarity criterion.
Evidently the similarity quickly drops off with increasing $\Delta_{\hat \varphi}$ and the higher the coherence the more dramatic the decline.
Consequently, the denoising performance suffers in hilly or mountainous terrain.
This effect is quite pronounced for bistatic TanDEM-X data due to their generally high coherence.

This analysis is not exclusive to the Kullback-Leibler divergence.
Similar arguments can be made for different similarity criteria, i.e., the one employed in~\cite{deledalle_nl-sar:_2015} and \cref{eq:pix_sim_likelihood}.

To combat the reduced denoising performance for terrain with significant height changes, we incorporated a linear fringe model as in~\cite{suo_local_fringe_2010} that accounted for the deterministic, topographic phase component when computing the similarities and the weighted mean.
Our approach is distantly related to~\cite{fedorov_affine_nonlocal_2017}, which employs affine transforms to find more similar patches.

For every pixel, the fringe compensation algorithm obtained an estimate of the fringe frequencies in azimuth and range $\mathbf{f} = {\left[ f_\textnormal{r}, f_\textnormal{az} \right]}^T$ using the 2D Fourier transform.
To circumvent abrupt changes of the fringe frequency estimates, we smoothed $\mathbf{f}$ with a Gaussian kernel.

Without loss of generality we can consider \cref{eq:kldivs} as a function of only the phase difference between two pixels

\begin{align}
    \delta^2_{\mathbf{x, y}}(\hat \varphi_\mathbf{x} - \hat \varphi_\mathbf{y}) \;.
\end{align}
The fringe compensation takes the fringe frequencies at $\mathbf{x}$ into account by changing the pixel similarity function to

\begin{align}
    \delta^2_\mathbf{x, y}(\hat \varphi_\mathbf{x} - (\hat \varphi_\mathbf{y} - {(\mathbf{x} - \mathbf{y})}^T \mathbf{f_x}) \bmod 2 \pi),
\end{align}
that is, we remove the phase component caused by the fringe frequency in azimuth and range.

The computation of the patch-wise weighted mean of the interferogram has to account for the phase model

\begin{align}
    \mathbf{\hat z}_\mathbf{x} = \sum\limits_{\mathbf{y} \in \partial_\mathbf{x}}  w_{\mathbf{x}, \mathbf{y}} \mathbf{z}_\mathbf{y} \cdot e^{- \iu {(\mathbf{x} - \mathbf{y})}^T \mathbf{F_x}} \;.
\label{eq:weighted_mean_patch_fringe_comp}
\end{align}
Here $\cdot$ denotes element-wise multiplication and $\mathbf{F} \in \mathds{R}^{2 \times p \times p}$ is a three dimensional tensor that contains all fringe frequencies of the pixels inside the $p \times p$ patch centered at $\mathbf{x}$.

\begin{figure}
\centering
    \makeatletter%
    \if@twocolumn%
      \newcommand{\subfigwidth}{0.95\columnwidth}
    \else
      \newcommand{\subfigwidth}{0.75\columnwidth}
    \fi
    \makeatother
  \begin{subfigure}[t]{\subfigwidth}
    \centering
    \includegraphics[width=\textwidth]{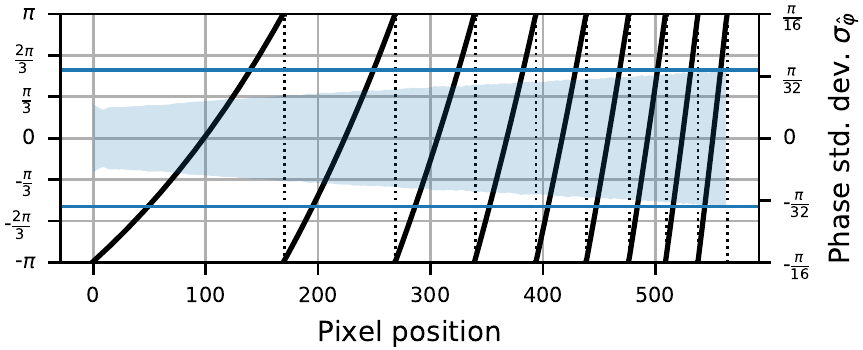}
    \caption{without fringe compensation}
  \end{subfigure}

  \begin{subfigure}[t]{\subfigwidth}
    \centering
    \includegraphics[width=\textwidth]{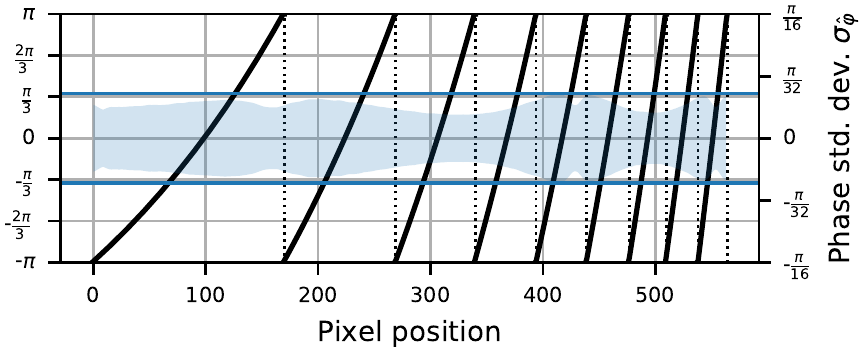}
    \caption{with fringe compensation}
  \end{subfigure}
  \caption{Standard deviation (shaded blue area) of NL-SWAG's estimate of a nonlinear phase profile (in black) with and without compensating for the fringe frequency. The maximum value of the standard deviation are marked with a horizontal blue line. If the filter does not account for the deterministic phase change inside the search window the denoising performance decreases substantially with increasing frequency.}%
  \label{fig:nonlinear_phase}
\end{figure}

\Cref{fig:nonlinear_phase} shows the effect that fringe frequency compensation has on the noise reduction.
Denoising of a nonlinear phase ramp with constantly increasing frequency was performed using NL-SWAG with and without fringe frequency compensation.
If the fringe frequency is not accounted for, the phase estimate's standard deviation increases steadily with increasing frequency.
With fringe frequency compensation, the standard deviation is limited.
Due to the discrete nature of the frequency estimation by fast Fourier transform in our implementation, the frequency was not perfectly estimated and the performance was not entirely frequency independent, which resulted in the wave-like pattern of the standard deviation.
A more sophisticated frequency estimation algorithm would certainly alleviate this problem.

As a final note, we would like to point out the difference between the fringe frequency compensation and the local phase heterogeneity-based adaptive patch size selection.
Both approaches address deterministic phase changes which can hamper the search for similar patches.
But whereas the fringe frequency compensation strictly deals with large-scale phase changes due to topography by a linear compensation, the role of the phase heterogeneity is more to take care of arbitrary small-scale phase changes, which would not necessarily be captured by a simple linear approximation.

\section{Experimental Results}%
\label{sec:experiments}

We compared NL-SWAG using simulations and real world data sets with existing nonlocal filters.
We used TanDEM-X bistatic strip map interferograms of three different test sites: Marseille, Munich, and Barcelona for the evaluation.
The most pertinent parameters are listed in \cref{tab:tdx_params}.
The experiments also substantiated our claim of creating a DEM close in quality to the HDEM specifications in~\cref{tab:demspec}.

\begin{table}
\caption{TanDEM-X strip map parameters of the test sites}
\begin{tabularx}{\columnwidth}{lXl}
\toprule
Parameter & Test site & Value \\
\midrule
Range bandwidth         & ---       & \SI{100}{\mega\hertz} \\
Ground range resolution & ---       & \SI{3}{\meter}\\
Azimuth resolution      & ---       & \SI{3}{\meter}\\
Polarization            & ---       & HH \\ 
Height of ambiguity     & Marseille & \SI{30}{\meter} \\ 
                        & Munich    & \SI{48}{\meter} \\ 
                        & Barcelona & \SI{48.5}{\meter} \\ 
\bottomrule
\end{tabularx}%
\label{tab:tdx_params}
\end{table}

In addition, the comparison included the result of a simple $5 \times 5$ Boxcar filter.
Boxcar filters, the dimensions of which depend on range resolution, incidence angle and imaging mode, are employed in DLR's integrated processor (ITP)~\cite{breit_itp_2010, fritz_itp_2011, rossi_tandemx_rawdem_2012} for generating the global TanDEM-X DEM\@.
For strip map data the dimensions of all employed Boxcar filters are close to $5 \times 5$ and their individual results will not be reported here.

We also analyzed NL-InSAR~\cite{deledalle_nl-insar:_2011}, the first nonlocal InSAR filter, where we set the search window size to $21 \times 21$, the patch size to $7 \times 7$ and used five iterations.
We deviated from the suggested ten iterations in the original publication as in our experience the changes in estimation accuracy are negligible after about four to five iterations.
Furthermore, the refinement provided by the iterations only resulted in improved detail preservation, which as we will show NL-InSAR already excels at, even with only five iterations.
Also, more iterations aggrevate the aforementioned terrace-like artifacts.

The second nonlocal filter in the comparison was NL-SAR~\cite{deledalle_nl-sar:_2015}.
NL-SAR adaptively selects the best parameters from a predefined set, which includes the patch size, search window size and the strength of the initial prefiltering step.
In our analysis, we used the same predefined set as in the original paper.

In all subsequent experiments concerning NL-SWAG, the search window size was set to $21 \times 21$, $h$ to $4$ in the first stage and to $2$ in the second stage.
The block size of the fringe estimation was $32 \times 32$ and the size of the discrete Fourier transform's was set to $64 \times 64$.
This zero padding increases the accuracy of the fringe estimation.

\subsection{Synthetic Data}

Assuming fully developed speckle, the correlated complex normal distributed pixels of two SLCs have the covariance matrix~\cite{goodman_multivar_complex_gaussian_1963}
\begin{align}
    \mathbf{C} = \begin{bmatrix}
        A^2 & A^2 \gamma e^{\iu \varphi} \\
        A^2 \gamma e^{- \iu \varphi} & A^2
\end{bmatrix}
\end{align}
where $A$ denotes the amplitude, $\varphi$ the interferometric phase and $\gamma$ the coherence.

Let $\mathbf{C} = \mathbf{L} \mathbf{L}^\dagger$ be the Cholesky decomposition of the covariance matrix $\mathbf{C}$, where $\dagger$ denotes conjugate transpose.
A multiplication with $\mathbf{L}$ transforms two independent complex normal distributed samples $r_1$ and $r_2$ of zero mean and unit variance
\begin{equation}
    \begin{bmatrix} u_1 \\ u_2 \end{bmatrix}
    = \mathbf{L}
    \begin{bmatrix} r_1 \\ r_2 \end{bmatrix}
    = A
    \begin{bmatrix} 1 & 0 \\ \gamma e^{- \iu \varphi} & \sqrt{1 - \gamma^2} \end{bmatrix}
    \begin{bmatrix} r_1 \\ r_2 \end{bmatrix}
\end{equation}
to samples with the desired correlation properties, amplitude and phase defined by the covariance matrix.

\begin{figure}
    \centering
    \makeatletter%
    \if@twocolumn%
      \newcommand{\figwidth}{\columnwidth}
    \else
      \newcommand{\figwidth}{0.75\textwidth}
    \fi
    \includegraphics[width=\figwidth]{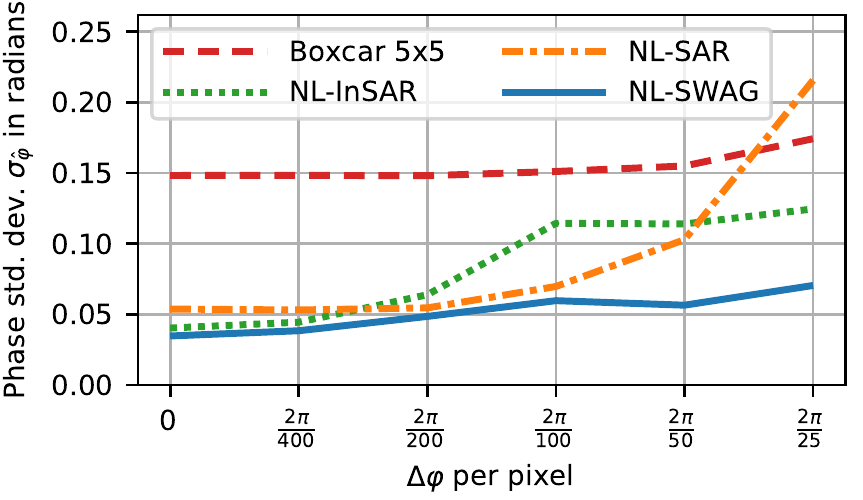}
    \caption{Standard deviation of the phase estimate as a function of a constant ramp's inclination. The steeper the incline the higher the standard deviation. The fringe frequency estimation of NL-SWAG alleviates this problem.}%
    \label{fig:slope_dependency}
\end{figure}

An analysis for the slope-dependent noise suppression was carried out by denoising phase ramps of different inclinations.
In the simulations, the intensity was constant for the whole slope and the coherence was set to 0.7.
\Cref{fig:slope_dependency} shows the standard deviation of the various filters' phase estimates for different inclinations, which is given as the phase change per pixel in radians.

The nonlocal filters are more sensitive to changes in inclination compared to the Boxcar filter as a result of their large search windows.
NL-InSAR and NL-SAR in particular, since they do not compensate for the deterministic phase component.
As mentioned earlier, the fringe estimation of NL-SWAG was not perfect due to the discrete nature of the fast Fourier transform used in the implementation and hence was still slope-dependent.
Overall, we can see that NL-SWAG provided an improvement of roughly a factor of three compared to the Boxcar estimate over all frequencies.

\begin{figure}
    \centering
    \makeatletter%
    \if@twocolumn%
      \newcommand{\subfigwidth}{0.49\columnwidth}
    \else
      \newcommand{\subfigwidth}{0.35\textwidth}
    \fi
    \makeatother
  \begin{subfigure}[t]{\subfigwidth}
    \centering
    \includegraphics[width=\textwidth]{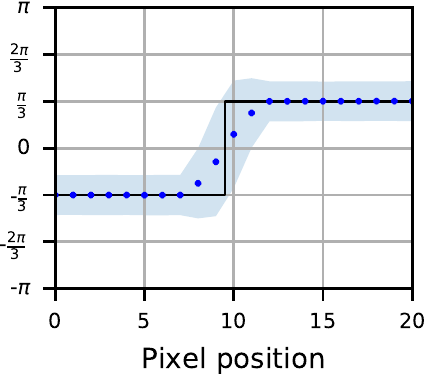}
    \caption{Boxcar $5\times5$}
  \end{subfigure}
  \begin{subfigure}[t]{\subfigwidth}
    \centering
    \includegraphics[width=\textwidth]{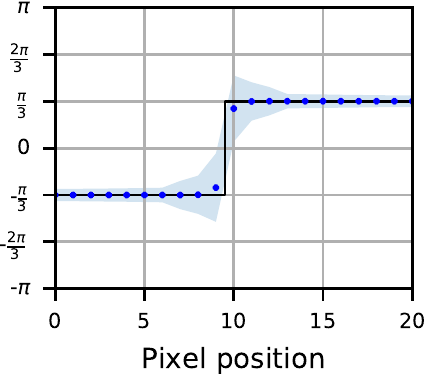}
    \caption{NL-InSAR}
  \end{subfigure}

  \begin{subfigure}[t]{\subfigwidth}
    \centering
    \includegraphics[width=\textwidth]{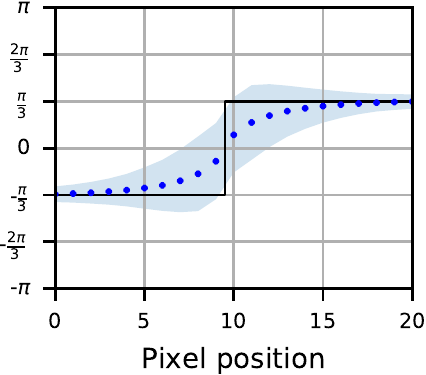}
    \caption{NL-SAR}
  \end{subfigure}
  \begin{subfigure}[t]{\subfigwidth}
    \centering
    \includegraphics[width=\textwidth]{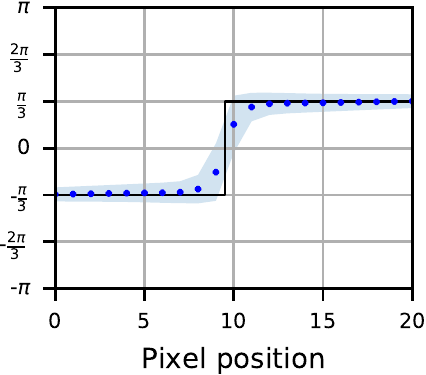}
    \caption{NL-SWAG}
  \end{subfigure}
  \caption{Expected value of a step function's phase estimate, constant amplitude and coherence of $0.7$. The shaded blue area delineates $\pm$ three times the estimate's standard deviation. We performed 10,000 simulations to obtain the statistics.}%
  \label{fig:phase_step}
\end{figure}

\begin{figure}
    \centering
    \makeatletter%
    \if@twocolumn%
      \newcommand{\subfigwidth}{0.49\columnwidth}
    \else
      \newcommand{\subfigwidth}{0.35\textwidth}
    \fi
    \makeatother
  \begin{subfigure}[t]{\subfigwidth}
    \centering
    \includegraphics[width=\textwidth]{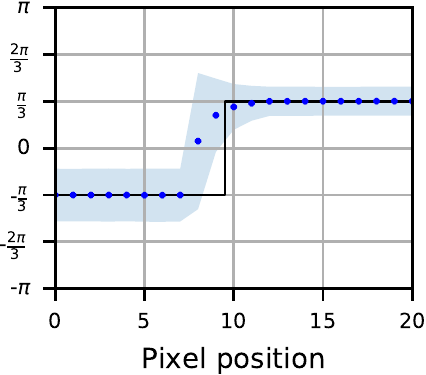}
    \caption{Boxcar $5\times5$}
  \end{subfigure}
  \begin{subfigure}[t]{\subfigwidth}
    \centering
    \includegraphics[width=\textwidth]{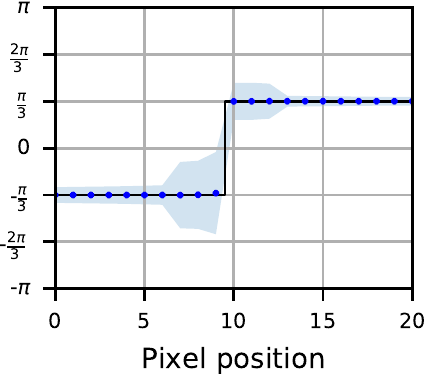}
    \caption{NL-InSAR}
  \end{subfigure}

  \begin{subfigure}[t]{\subfigwidth}
    \centering
    \includegraphics[width=\textwidth]{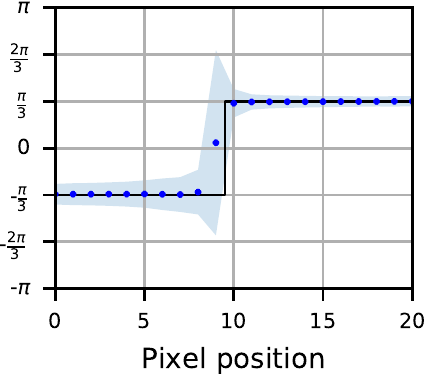}
    \caption{NL-SAR}
  \end{subfigure}
  \begin{subfigure}[t]{\subfigwidth}
    \centering
    \includegraphics[width=\textwidth]{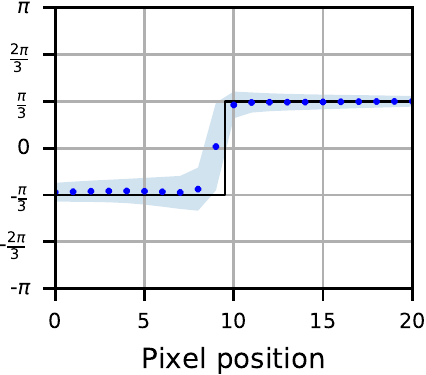}
    \caption{NL-SWAG}
  \end{subfigure}
  \caption{Estimated phase of a step function with a step in coherence from 0.6 to 0.8 and a intensity jump of 6\si{\decibel}. The additional change in intensity compared to \cref{fig:phase_step} helped the nonlocal filters to preserve the edge.}%
  \label{fig:phase_intensity_coherence_step}
\end{figure}

\cref{fig:phase_step} and \cref{fig:phase_intensity_coherence_step} give an impression of the resolution preservation capabilities of the various filters.
Both figures are the result of Monte-Carlo simulations with 10,000 repetitions when estimating a phase jump from $-\tfrac{\pi}{3}$ to $\tfrac{\pi}{3}$.
The expected values are plotted as blue dots and their standard deviations as shaded blue areas.
In \cref{fig:phase_step}, intensity and coherence are constant, with coherence having a value of 0.7, whereas in \cref{fig:phase_intensity_coherence_step} coherence increases from 0.6 to 0.8 and the intensity difference is $\SI{6}{\decibel}$.

\cref{fig:phase_step} shows that the Boxcar filter's result exhibited the expected smoothing.
Both NL-InSAR and NL-SWAG were unable to perfectly preserve the edge but fared much better than NL-SAR\@.
The reason for NL-SAR's poor performance is that NL-SAR initially produces an intentionally oversmoothed result and then applies a bias-reduction step based on terrain heterogeneity.
This heterogeneity test, however, only considers the intensity and therefore breaks down in this particular case, where only the phase changes.

The situation changed when the phase jump was accompanied by an intensity jump as in \cref{fig:phase_intensity_coherence_step}.
The intensity change aids nonlocal filters in discriminating between similar pixels, resulting in sharper transitions.
The benefit of setting the patch size adaptively is highlighted by NL-SAR and NL-SWAG, which do not exhibit a halo of high variance at the discontinuity.
We could deduce that the rare patch effect was indeed the cause of this performance degradation, as the width of the halo for NL-InSAR was equal to the employed patch size minus one.
All patches in this area included the edge and consequently suffered from the rare patch effect.
NL-SWAG additionally benefited by the aggregation step, which further reduced the variance along the edge.
Even with these measures in place, we could still see that the variance is increased near the edge.

\begin{figure*}
\centering
\includegraphics[width=\textwidth]{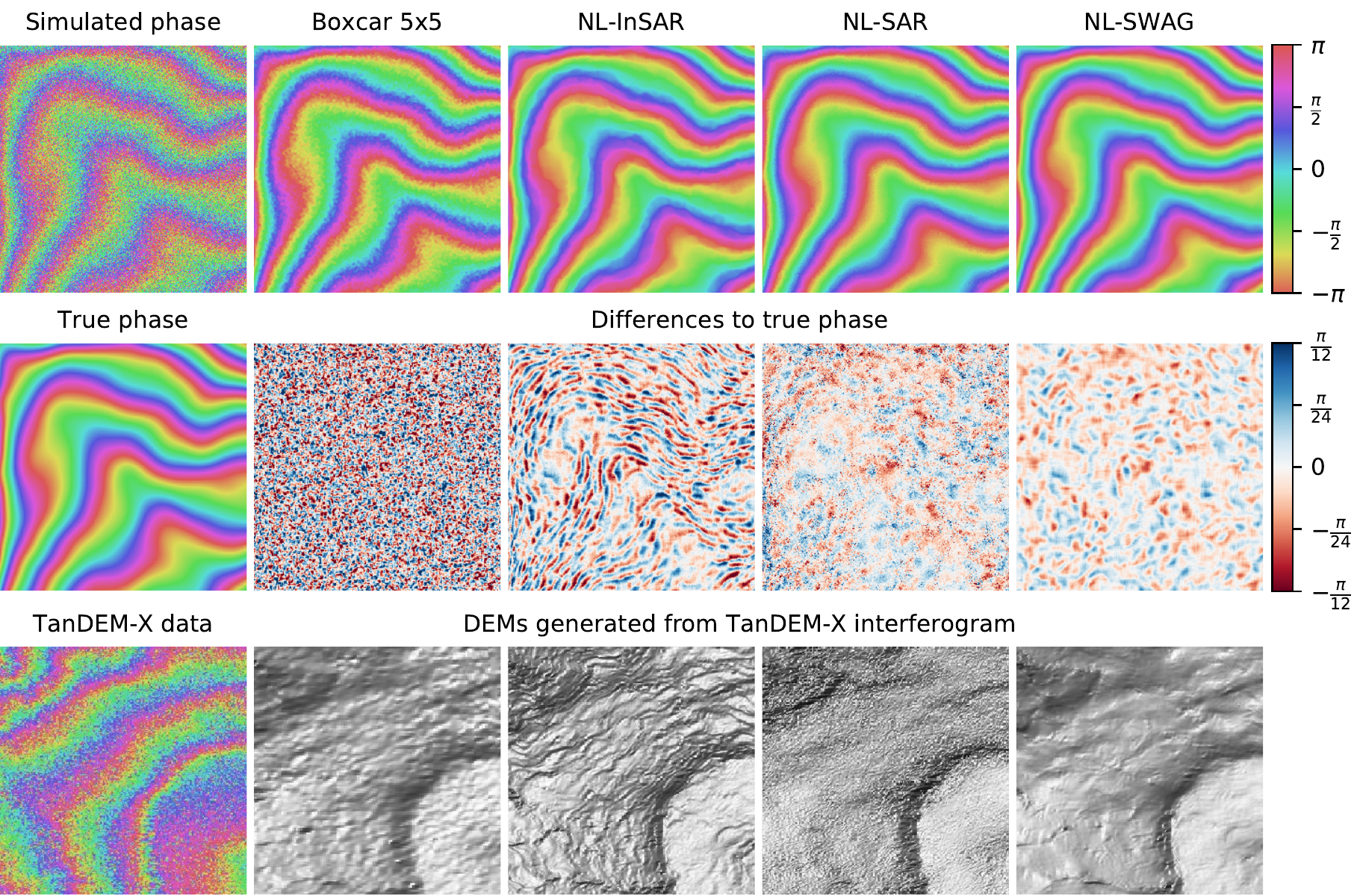}
\caption{Phase estimates of several filters for a synthetically-generated interferogram and their differences compared to the true phase are shown together with the noisy interferogram (the coherence was set to $0.7$) and the true phase in the first two rows.
The last row shows a comparable TanDEM-X strip map interferogram and the shaded relief of DEMs generated by the corresponding filter.
The phase estimate of NL-InSAR shows a distinct staircase-like pattern, which is also clearly visible in the shaded relief plot.
All nonlocal filters suppress the high-frequency component of the noise but produce low frequency undulations in the estimate.}%
\label{fig:fractal}
\end{figure*}

To illustrate the propensity of the filters to produce the earlier introduced terrace-like features and other biasing artifacts, we simulated a noisy interferogram from a synthetic terrain created by the diamond-square algorithm~\cite{fournier_diamond_square_1982}.
\cref{fig:fractal} shows in the top row the simulated noisy interferogram with a constant coherence value of $0.7$ and the filters' denoised results.
The second row shows the true simulated phase and its difference compared to the filter output.

We also include a TanDEM-X interferogram whose phase resembles the simulation in our analysis to exemplify how these filtering characteristics affect real data, which is shown in the last row together with shaded reliefs of DEMs generated by the various filters.

For NL-InSAR, a distinct pattern was visible in the difference plot which would manifest as terrace-like artifacts in a generated DEM\@.
Indeed, the DEM produced by NL-InSAR from real data also exhibited similar patterns.
Visually, we could asses that the overall noise level of all nonlocal filters was lower compared to the Boxcar filter, especially in regions where the fringe frequency was low.
The difference plots also show that nonlocal filters suppressed the high-frequency component of the noise but created slowly varying undulations of spatially correlated noise.

\begin{figure}
    \centering
    \makeatletter%
    \if@twocolumn%
      \newcommand{\figwidth}{\columnwidth}
    \else
      \newcommand{\figwidth}{0.85\textwidth}
    \fi
    \makeatother
    \includegraphics[width=\figwidth]{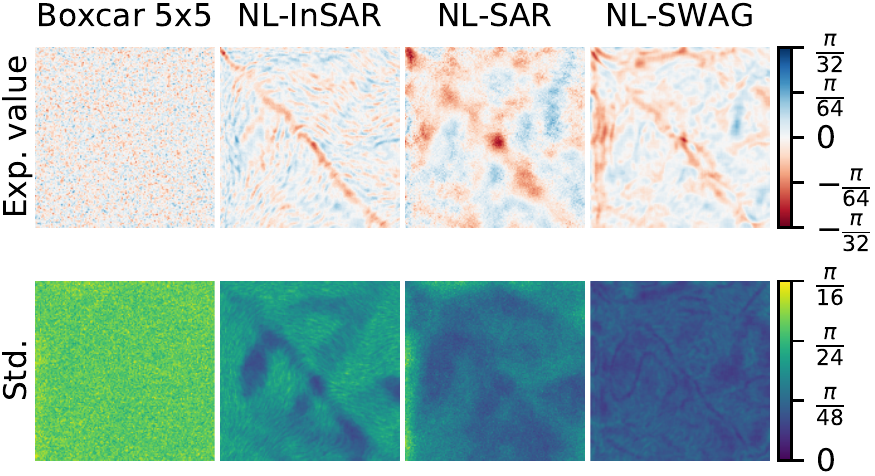}
    \caption{Expected values (top) and standard deviation (bottom) for a Monte-Carlo simulation of the simulated phase in \Cref{fig:fractal}. Minor biases are present in the phase estimates. The slope dependent denoising performance of nonlocal filters is evident in the standard deviation plots.}%
    \label{fig:fractal_monte_carlo}
\end{figure}

To further shed light on some of the mechanisms of nonlocal filters, \cref{fig:fractal_monte_carlo} shows the expected value and standard deviation of a Monte-Carlo simulation's phase estimate for the experiment with synthetic data in \cref{fig:fractal}.
All nonlocal filters biased the estimate along the ridge at the interferogram's diagonal.
In general, nonlocal filters have a higher propensity to bias the estimate due to their comparatively large search windows.
The standard deviation plots show the fringe-frequency dependent noise suppression of NL-InSAR and NL-SAR\@.
NL-SWAG was much less affected by this aspect, although it was also not completely immune as noted earlier.
\cref{tab:fractal_std} lists the mean standard deviations and the average equivalent number of looks, rounded to the nearest integer, over the whole image and all simulation runs.
In accordance with our previous experiments, it was considerably lower for nonlocal filters.
Contrasting \cref{tab:fractal_std} with \cref{tab:demspec} reveals that NL-SWAG would fulfill the noise reduction by a factor of 2.5, which is required for the production of a DEM according to the HDEM specifications.

\begin{table}
\caption{Standard deviation in radians and average equivalent number of looks, rounded to the nearest integer, for the Monte-Carlo simulation in \cref{fig:fractal_monte_carlo}}
\begin{tabularx}{\columnwidth}{lllll}
\toprule
& Boxcar $5 \times 5$ & NL-InSAR & NL-SAR & NL-SWAG \\
\midrule
$\sigma_{\hat \varphi}$ in \si{\radian} & 0.1482 & 0.0969     & 0.0768   & 0.0537 \\
Number of looks & 25  & 58    & 93   & 190 \\
\bottomrule
\end{tabularx}%
\label{tab:fractal_std}
\end{table}

\subsection{Real Data}

Experiments on TanDEM-X bistatic strip map interferograms were carried out for three test sites that were chosen to showcase the previously described qualities and phenomena when using nonlocal filters and NL-SWAG in particular for DEM generation.
The interferograms from the test sites were processed with DLR's ITP, and the aforementioned nonlocal filters were used in lieu of the default Boxcar filter.

\begin{figure}[ht!]
  \centering
    \makeatletter%
    \if@twocolumn%
      \newcommand{\subfigwidth}{0.49\columnwidth}
    \else
      \newcommand{\subfigwidth}{0.28\columnwidth}
    \fi
  \begin{subfigure}[t]{\subfigwidth}
    \centering
    \includegraphics[width=\textwidth]{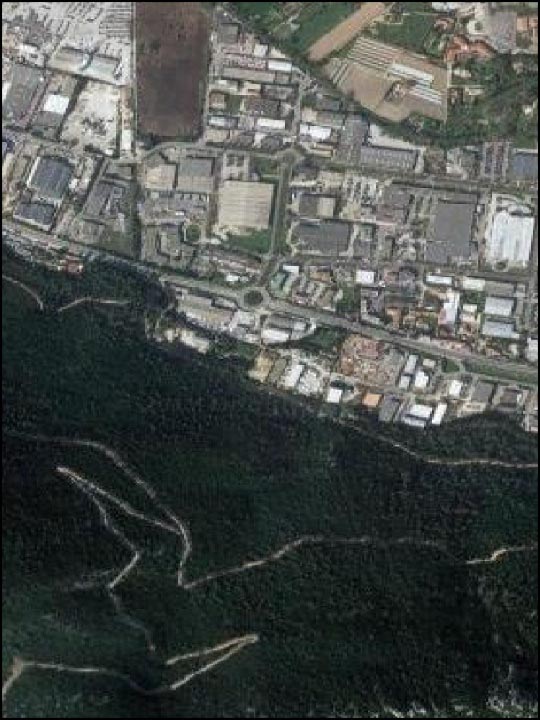}
  \caption{Optical, \textcopyright~Google}
  \end{subfigure}
  \begin{subfigure}[t]{\subfigwidth}
    \centering
    \includegraphics[width=\textwidth]{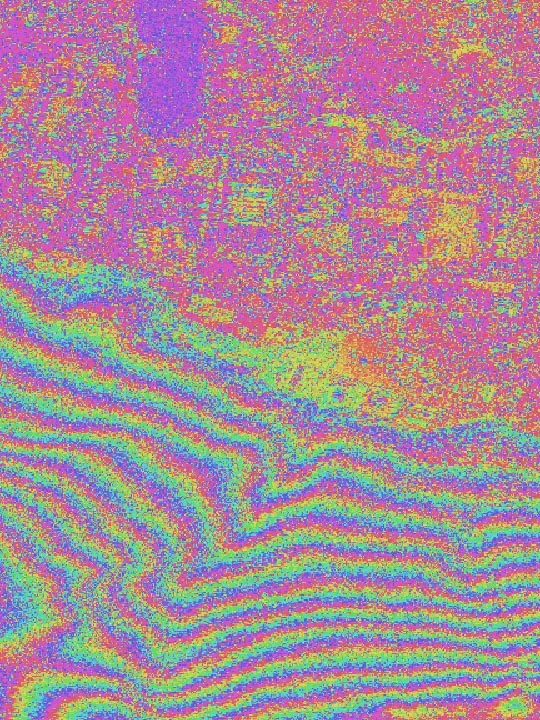}
  \caption{Unfiltered phase}
  \end{subfigure}

  \begin{subfigure}[t]{\subfigwidth}
    \centering
    \includegraphics[width=\textwidth]{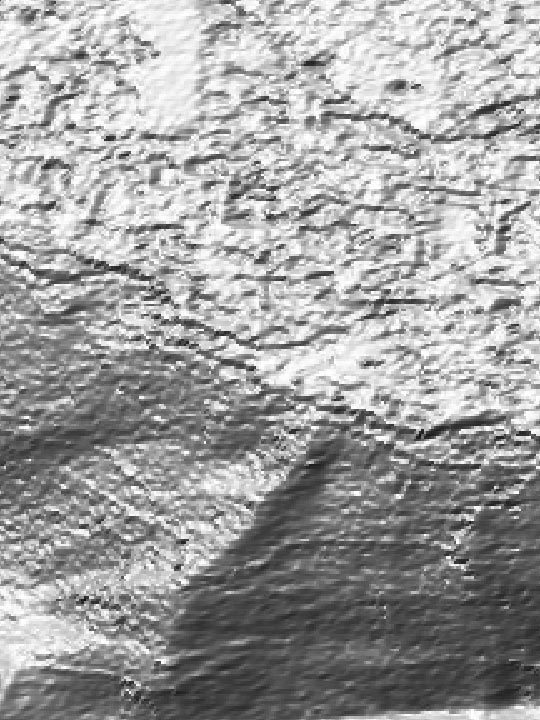}
    \caption{Boxcar $5\times5$}
  \end{subfigure}
  \begin{subfigure}[t]{\subfigwidth}
    \centering
    \includegraphics[width=\textwidth]{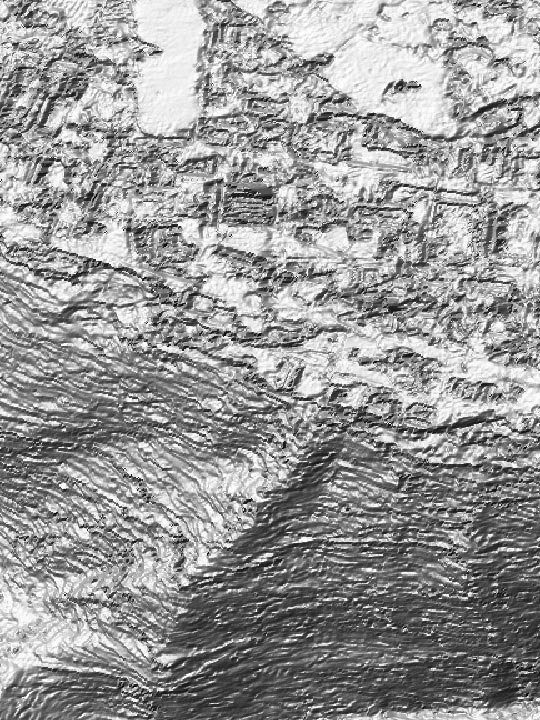}
    \caption{NL-InSAR}
  \end{subfigure}

  \begin{subfigure}[t]{\subfigwidth}
    \centering
    \includegraphics[width=\textwidth]{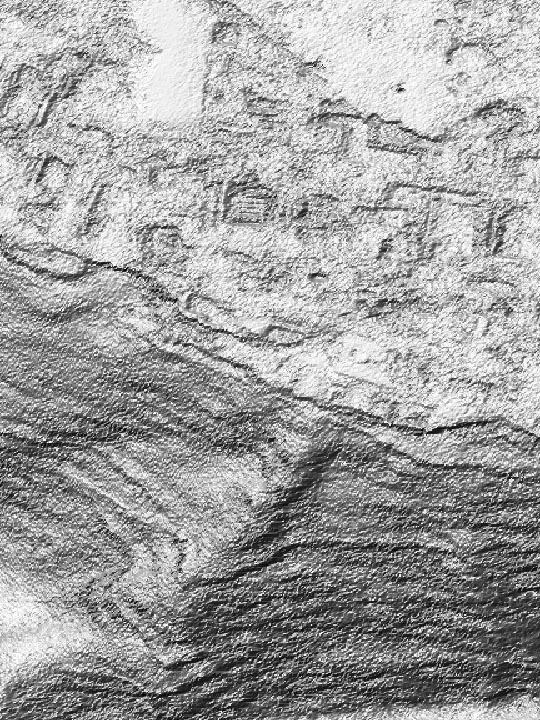}
    \caption{NL-SAR}
  \end{subfigure}
  \begin{subfigure}[t]{\subfigwidth}
    \centering
    \includegraphics[width=\textwidth]{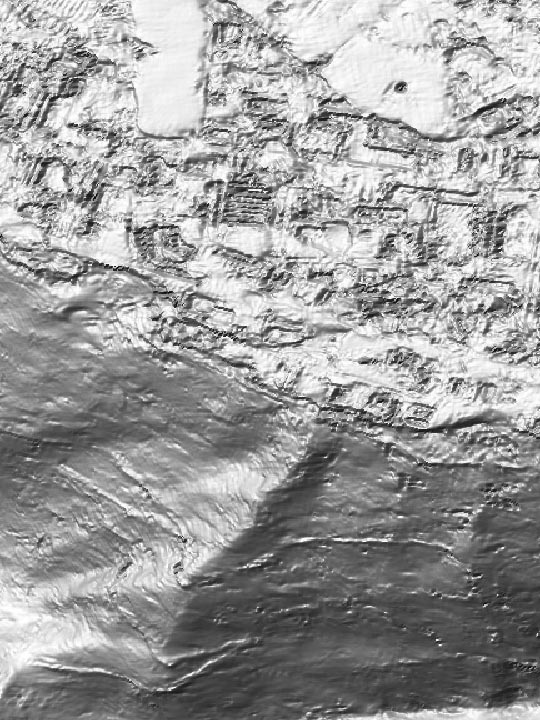}
    \caption{NL-SWAG}
  \end{subfigure}
  \caption{Shaded reliefs of DEMs generated with the various filters.
  The nonlocal filters improved the resolution and noise level compared to the Boxcar estimate.
  NL-InSAR suffered from the rare patch effect near structures due to its fixed patch size.}%
  \label{fig:dems_marseille}
\end{figure}

The first test area was an industrial site near the French city of Marseille and it provided a visual impression of the performance increase that could be expected with nonlocal filters.
\cref{fig:dems_marseille} shows shaded reliefs of the generated DEMs, an optical image for better interpretation and a plot of the unfiltered phase.
The resolution of the DEMs produced with the nonlocal filters was \SI{6}{\meter} for longitude and latitude.
The DEM generated using the $5 \times 5$ Boxcar filter had a resolution of \SI{12}{\meter}, the default configuration for DLR's RawDEM\@.
In the global TanDEM-X DEM processing chain, several RawDEMs are later combined to generate the final DEM product.

The higher level of details visible in the nonlocal DEMs is evident, as is the improved noise reduction for agricultural fields and the hill to the south.
NL-InSAR produced clearly discernible terraces for the hill, a result of the staircasing effect.
The road in the lower half of the image serves as an example for what kind of details can be preserved by the proposed filter.

Also noticeable are noisy artifacts near buildings for NL-InSAR at the industrial site, a consequence of the rare patch effect, which is avoided by NL-SAR and NL-SWAG\@.
NL-SAR, however, tends to oversmooth some details, so that, for example, the road in the lower part of the test site is hardly distinguishable from its surrounding.

\begin{figure}[htb]
  \centering
    \makeatletter%
    \if@twocolumn%
      \newcommand{\subfigwidth}{0.49\columnwidth}
    \else
      \newcommand{\subfigwidth}{0.28\columnwidth}
    \fi
  \begin{subfigure}[t]{\subfigwidth}
    \centering
    \includegraphics[width=\textwidth]{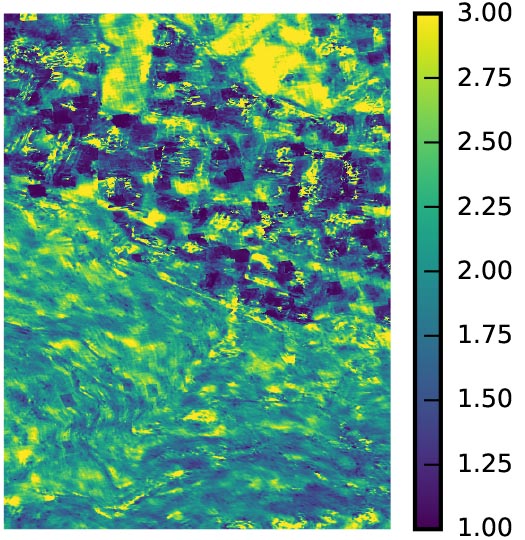}
    \caption{Sigma of Gaussian windows}
  \end{subfigure}
  \begin{subfigure}[t]{\subfigwidth}
    \centering
    \includegraphics[width=\textwidth]{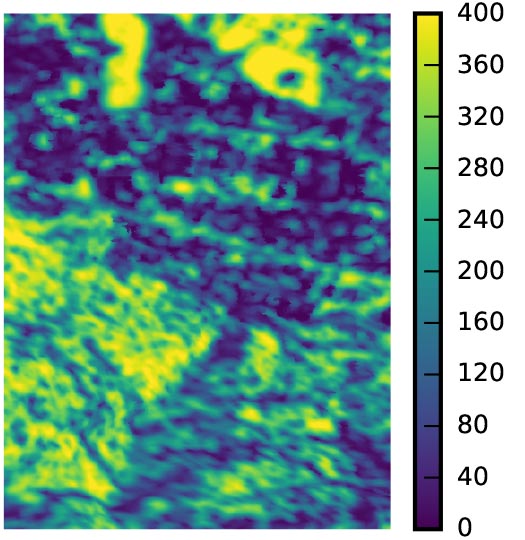}
  \caption{Equivalent number of looks}
  \end{subfigure}
  \caption{Width of the Gaussian windows used for computing the patch similarities and the equivalent number of looks for the test site from \cref{fig:dems_marseille}.}%
  \label{fig:dems_marseille_app}
\end{figure}

\cref{fig:dems_marseille_app} sheds some more light on NL-SWAG's filtering characteristics.
It shows the employed width of the Gaussian window used for computing the patch similarities and the final equivalent number of looks after the aggregation step.
Both show that homogeneous areas benefit from wide Gaussian windows, resulting in accurate patch similarity estimates, and a large number of similar pixels within the search window, leading to low-noise estimates.
The reverse is true for the industrial site, where narrow Gaussian windows were employed, due to the region's heterogeneity.
This heterogeneity was also the cause of only a comparatively low number of looks.
The impact that the fringe frequency estimation and compensation had on the estimate could be inferred from the equivalent number of looks for the hilly terrain to the south, which was virtually unaffected by the trend of the phase.

\begin{figure}[htbp]
  \centering
    \makeatletter%
    \if@twocolumn%
      \newcommand{\subfigwidth}{0.48\columnwidth}
    \else
      \newcommand{\subfigwidth}{0.33\columnwidth}
    \fi
  \begin{subfigure}[t]{\subfigwidth}
    \centering
    \includegraphics[width=\textwidth]{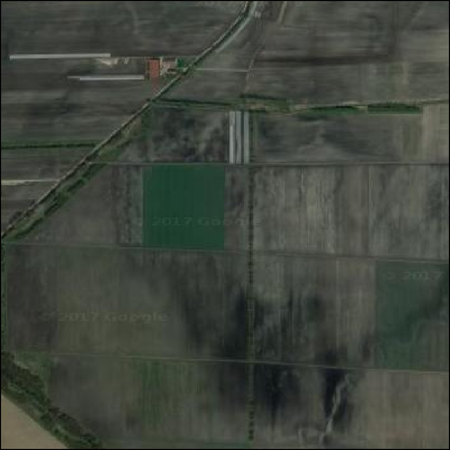}
  \caption{Optical, \textcopyright~Google}
  \end{subfigure}
  \begin{subfigure}[t]{\subfigwidth}
    \centering
    \includegraphics[width=\textwidth]{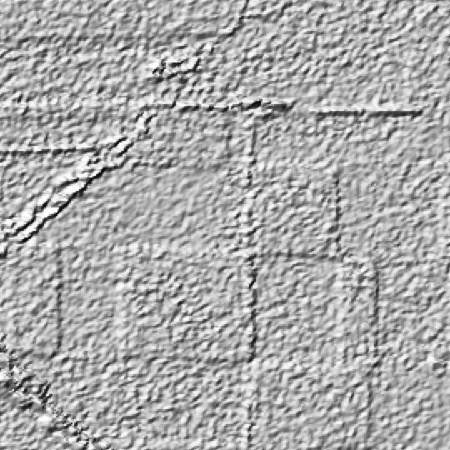}
    \caption{Boxcar $5\times5$}
  \end{subfigure}

  \begin{subfigure}[t]{\subfigwidth}
    \centering
    \includegraphics[width=\textwidth]{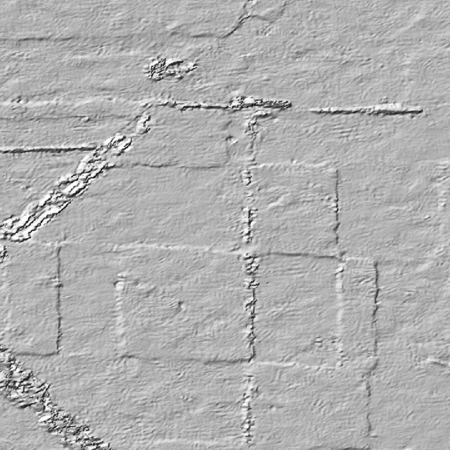}
    \caption{NL-InSAR}
  \end{subfigure}
  \begin{subfigure}[t]{\subfigwidth}
    \centering
    \includegraphics[width=\textwidth]{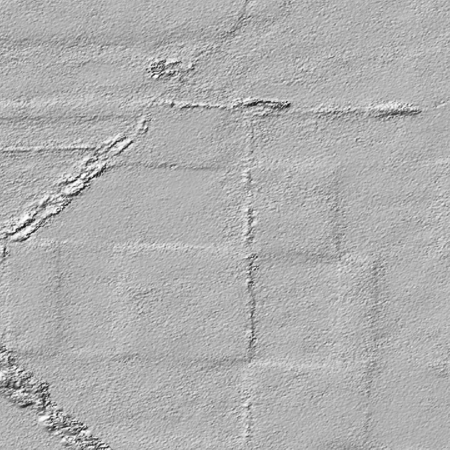}
    \caption{NL-SAR}
  \end{subfigure}

  \begin{subfigure}[t]{\subfigwidth}
    \centering
    \includegraphics[width=\textwidth]{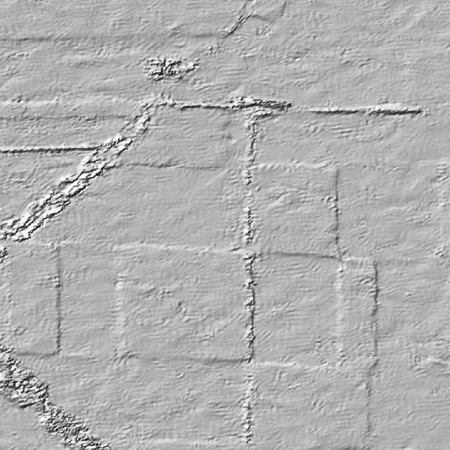}
    \caption{NL-SWAG $h=\frac{1}{2}$}
  \end{subfigure}
  \begin{subfigure}[t]{\subfigwidth}
    \centering
    \includegraphics[width=\textwidth]{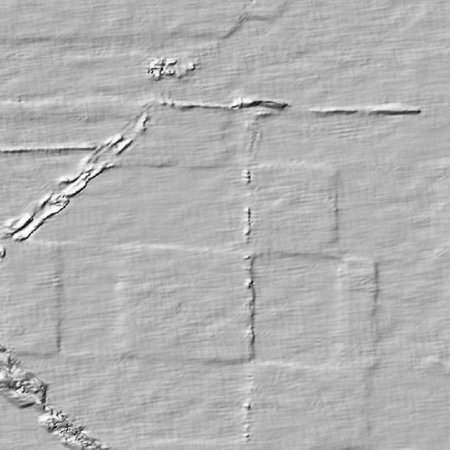}
    \caption{NL-SWAG $h=2$}
  \end{subfigure}
  \caption{Shaded reliefs of DEMs of an agricultural site. Clearly visible are height changes between fields. The bottom row shows the effect that changing $h$ in the second stage has on detail preservation and noise reduction.}%
  \label{fig:dems_munich}
\end{figure}

As a clearer example of detail preservation, \cref{fig:dems_munich} shows DEMs for an agricultural area near Munich, Germany.
The resolution was the same as in the previous example: \SI{6}{\meter} for the nonlocal DEMs and \SI{12}{\meter} for the Boxcar filter.
The data were acquired on August 19, 2011 when some of the fields had already been harvested so the outlines of different fields are clearly discernible, as electromagnetic waves in X-Band only marginally penetrate vegetation~\cite{rossi_paddy_rice_2015}.
The shaded reliefs confirmed our simulation results in \cref{fig:phase_step} and \cref{fig:phase_intensity_coherence_step} that NL-InSAR provided the best result for this particular scenario, as it favors piecewise constant solutions and sharp edges.
But this propensity was also the source of the highly unwelcomed staircasing for regions with a more interesting topographic profile.

We can also see the effect that a change of $h$ has on the filtering result.
A lower value of $h$ produced a sharper transition at the edges of the field but reduced denoising in flat terrain.

\begin{figure*}[htbp]
    \includegraphics[width=\textwidth]{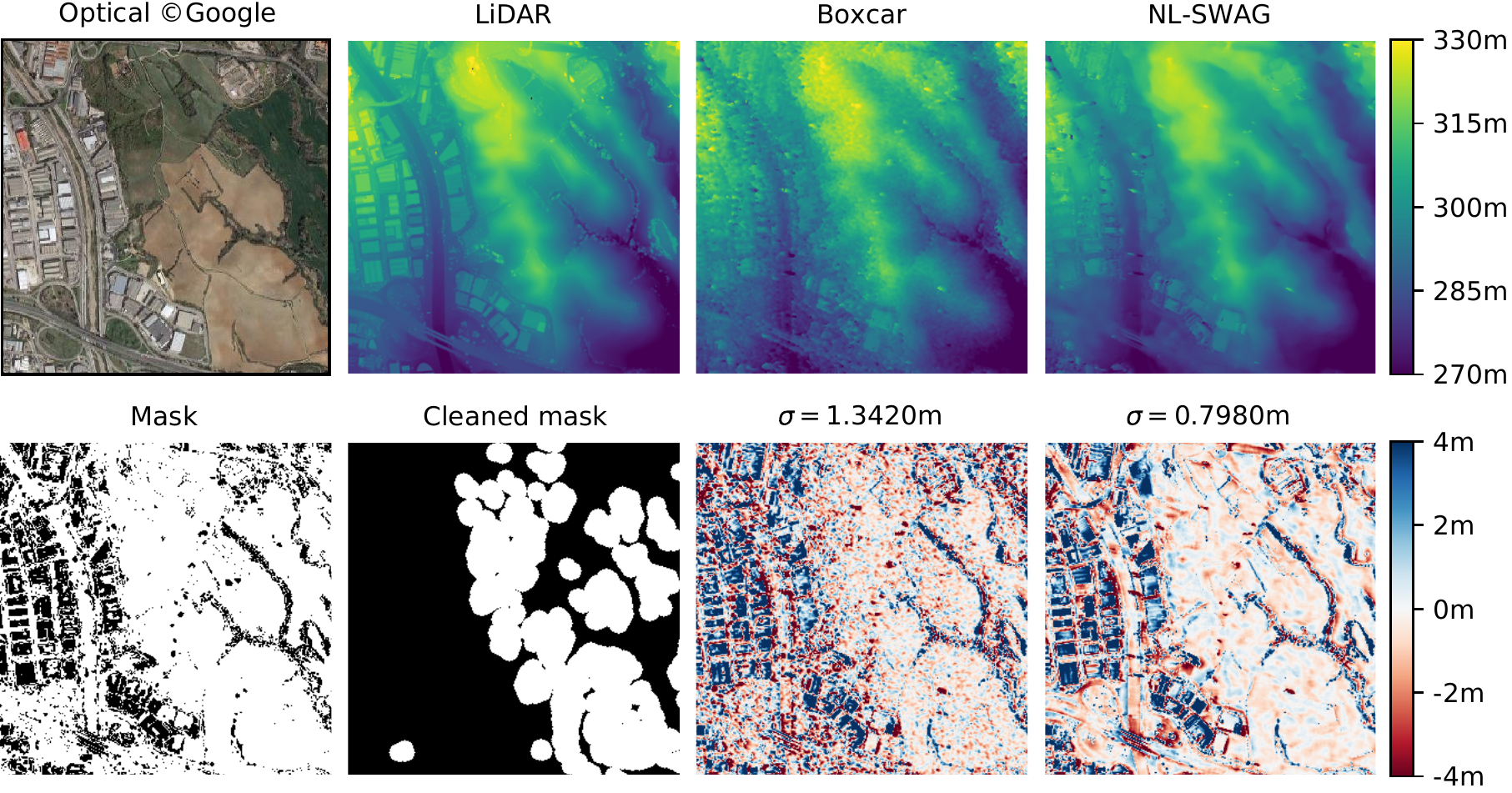}
    \caption{DEMs generated by NL-SWAG and a $5 \times 5$ Boxcar filter from a TanDEM-X interferogram are compared to a LiDAR DEM\@.
    The bottom row shows the height differences compared to the LiDAR DEM\@.
    For the masked-out area, standard deviations were computed for the height differences.}%
    \label{fig:lidar_terrassa}
\end{figure*}

As a last example we compared NL-SWAG to a high-resolution LiDAR DEM, which served as a gold standard for our analysis.
The test site was the town Terrassa close to Barcelona in Spain.
The top row in \cref{fig:lidar_terrassa} shows an optical image from Google maps and the LiDAR DEM with \SI{5}{\meter} spacing plus DEMs generated from a single TanDEM-X interferogram by a $5 \times 5$ Boxcar filter and our proposed method.
The DEMs were resampled to the grid of the LiDAR DEM\@.
As LiDAR and SAR have fundamentally different imaging geometries and properties, we tried to remove areas with systematic errors, such as urban areas suffering from layover and shadowing or vegetation, where the LiDAR's last returns differed from the scattered wave's phase center at X-Band.
In order to do so, we compared the LiDAR DEM to the global TanDEM-X DEM and excluded points with a height difference larger than \SI{2}{\meter}.
The result is depicted in the bottom row of \cref{fig:lidar_terrassa} and a cleaned mask, using morphological operations, right next to it.
The height differences are the remaining two pictures annotated with the standard deviation of the height difference computed over the masked area.

This experiment had several noteworthy results.
As expected, the SAR DEMs differed substantially from the LiDAR DEM for buildings, and the height values were unusable.
However the SAR DEMs could still be used to detect buildings as the test site near Marseille (\cref{fig:dems_marseille}) showed as well.
On the masked-out, moderately hilly, homogeneous terrain, NL-SWAG improved the noise level roughly by a factor of $\SI{1.3420}{\meter} / \SI{0.7980}{\meter} \approx 1.6817 $ almost equivalent to a filter with three times as many looks, which is, however, insufficient for completely fulfilling the requirements in \cref{tab:demspec}.

At first glance, this improvement in noise reduction contradicted our findings reported in \cref{tab:fractal_std}.
We could exclude systematic height differences due to the different physical properties of LiDAR and SAR as the penetration depth of electromagnetic waves at X-Band is negligible as an error source~\cite{nolan_penetration_depth_2003}.
Coregistration errors of the LiDAR and SAR DEMs might also be a contributing factor for height differences but for the moderately hilly terrain they would only play a minor role.
Such error sources would equally increase the difference compared to the LiDAR DEM, leading to a misrepresented noise level reduction.
The true reason for this discrepancy is the resampling from approximately $\SI{3}{\meter}$ pixel spacing in range and azimuth to the $\SI{5}{\meter}$ LiDAR pixel spacing, which essentially increased the footprint of the Boxcar filter.
For NL-SWAG, this effect was imperceptible due to its comparatively large search window.

\section{Discussion}%
\label{sec:discussion}

The initial goal of our investigation was to ascertain whether nonlocal filters were suitable for generating a DEM close to the HDEM standard (see \cref{tab:demspec}) from the globally available TanDEM-X data.
In the following paragraphs, we will detail how the proposed filter held up to these challenges.

All conducted experiments confirmed that nonlocal filters were able to deliver a vastly improved noise reduction over the exemplary local Boxcar filter.
The reason is that, due to their large search windows, nonlocal filters found a multitude of pixels for the averaging process, even for comparatively heterogeneous terrain.
To further quantify this improvement: For the experiments on synthetic data (\cref{fig:slope_dependency} and \cref{tab:fractal_std}) the standard deviation was lower by a factor of three and for the real data set of \cref{fig:lidar_terrassa} on moderately complex terrain it was still reduced by a factor of approximately $1.7$.
Relating this to the level of noise reduction we aimed for in \cref{tab:demspec}, our filter fell short of reaching the target of $2.5$ roughly by a factor of $\sqrt{2}$.
Depending on the type of terrain, this might still be sufficient to obtain a DEM that fulfills the requirements of the HDEM, as already the globally available TanDEM-X DEM often overfulfills its accuracy requirements.
In any case, having twice as many acquisitions available would also satisfy the specification.

Our proposed filter implemented several techniques to reach this level of noise reduction.
It reduced the detrimental effect of topography by its fringe frequency compensation accounting for the deterministic topographic phase component, as evidenced by \cref{fig:slope_dependency} and \cref{fig:fractal_monte_carlo}.
Furthermore, even on flat, homogeneous terrain, the high inherent noise level of InSAR hampered denoising, which was countered by the two-step approach.

\cref{fig:fractal_monte_carlo} shows that nonlocal filters bias the estimate for nonlinear phase profiles.
The bias is limited by approximately $\pm \tfrac{\pi}{100}$.
With a height of ambiguity of \SI{40}{\meter}, which is a typical value for TanDEM-X interferograms, this translates to deterministic height errors of $\pm$ \SI{20}{\centi\meter}, well within the HDEM specifications.

We also highlighted that for nonlocal filters it is far easier to denoise homogeneous terrain than heterogeneous targets, as more similar pixels are found.
Nonetheless, nonlocal filters were well suited for preserving heterogeneous targets as shown by the simulation results in \cref{fig:phase_step} and \cref{fig:phase_intensity_coherence_step}, where the adaptive patch size and the aggregation step played a significant role to avoid the rare patch effect near the edge.

For filtering SAR interferograms of urban areas or terrain with man-made structures, nonlocal filters were especially appealing as such heterogeneous targets exhibit a very high radar cross-section compared to their surroundings.
These high intensity variations aid nonlocal filters to preserve details as their weight maps are more discriminant. 
The gain in resolution, compared to simple boxcar averaging, is evident for real data in \cref{fig:dems_marseille} and \cref{fig:dems_munich}.

It wold also be rather straightforward to extend existing nonlocal filters with the proposed modifications.
The fringe compensation requires only a minor adaption of the similarity criterion.
Changing the patch size adaptively is an isolated modification, which could also be performed based on the intensity heterogeneity criterion derived in~\cite{lee_speckle_analysis_1981}, for example, in a nonlocal SAR despeckling filter.
The aggregation step is an extension of the pixelwise weighted mean and can be treated separately from all other adjustments.

\section{Conclusion}%
\label{sec:conclusion}

We showed that applying existing nonlocal filters led to artifacts when generating DEMs.
Our analysis highlighted the mechanisms behind the encountered phenomena, like the topographic phase component and the myriad types of terrain and settings, from agricultural fields to city landscapes, in which InSAR filters have to operate.
The proposed filter addressed these issues by accounting for the deterministic fringe frequency and setting its filtering parameters adaptively.
We demonstrated the effectiveness of these measures which resulted in a comparable noise reduction and detail preservation compared to other nonlocal InSAR filters without any of the undesired properties.
The derived DEMs also far surpassed the RawDEMs produced with the existing global TanDEM-X processing chain, which relies on conventional boxcar multilooking.

We will further evaluate the proposed method on a wider array of real data, which will also highlight some of the characteristics of SAR compared to LiDAR for generating DEMs.
Such an extensive evaluation is essential for considering nonlocal filters as a total replacement for the boxcar filter in the TanDEM-X processing chain.

Promising paths for future research include exploiting spatial redundancies within a patch as is the case with SAR-BM3D~\cite{parrilli_nonlocal_2012} and also taking into account the slope-dependent reflectivity when computing similarities.
The robustness of the filter could be increased by also setting the search window dimensions adaptively depending on the local scene heterogeneity, which could also be achieved by designing a weighting kernel with thresholding.
Furthermore, the proposed filter only relies on the interferometric phase to classify scenes as heterogeneous, also taking the intensity into account might provide more accurate estimates, especially in urban areas.

\section*{Acknowledgment}

The authors would like to thank the anonymous reviewers for their valuable comments.
They would also like to mention the enlightening discussions they had with their colleagues Thomas Fritz, Helko Breit and Michael Eineder from DLR, as well as Michael Schmitt of the Technical University of Munich.


\bibliographystyle{IEEEtran}
\bibliography{references}

\begin{thebibliography}{10}
\providecommand{\url}[1]{#1}
\csname url@samestyle\endcsname
\providecommand{\newblock}{\relax}
\providecommand{\bibinfo}[2]{#2}
\providecommand{\BIBentrySTDinterwordspacing}{\spaceskip=0pt\relax}
\providecommand{\BIBentryALTinterwordstretchfactor}{4}
\providecommand{\BIBentryALTinterwordspacing}{\spaceskip=\fontdimen2\font plus
\BIBentryALTinterwordstretchfactor\fontdimen3\font minus
  \fontdimen4\font\relax}
\providecommand{\BIBforeignlanguage}[2]{{%
\expandafter\ifx\csname l@#1\endcsname\relax
\typeout{** WARNING: IEEEtran.bst: No hyphenation pattern has been}%
\typeout{** loaded for the language `#1'. Using the pattern for}%
\typeout{** the default language instead.}%
\else
\language=\csname l@#1\endcsname
\fi
#2}}
\providecommand{\BIBdecl}{\relax}
\BIBdecl

\bibitem{lee_sigma_filter_1983}
J.~S. Lee, ``A simple speckle smoothing algorithm for synthetic aperture radar
  images,'' \emph{IEEE Transactions on Systems, Man, and Cybernetics}, vol.
  SMC-13, no.~1, pp. 85--89, 01 1983.

\bibitem{lee_improved_sigma_filter_2009}
J.~S. Lee, J.~H. Wen, T.~L. Ainsworth, K.~S. Chen, and A.~J. Chen, ``Improved
  sigma filter for speckle filtering of sar imagery,'' \emph{IEEE Transactions
  on Geoscience and Remote Sensing}, vol.~47, no.~1, pp. 202--213, 01 2009.

\bibitem{lee_polsar_speckle_filtering_extended_sigma_2015}
J.~S. Lee, T.~L. Ainsworth, Y.~Wang, and K.~S. Chen, ``Polarimetric sar speckle
  filtering and the extended sigma filter,'' \emph{IEEE Transactions on
  Geoscience and Remote Sensing}, vol.~53, no.~3, pp. 1150--1160, 03 2015.

\bibitem{buades_non-local_2005}
A.~Buades, B.~Coll, and J.-M. Morel, ``A non-local algorithm for image
  denoising,'' in \emph{{IEEE} {Computer} {Society} {Conference} on {Computer}
  {Vision} and {Pattern} {Recognition}, 2005. {CVPR} 2005}, vol.~2, Jun. 2005,
  pp. 60--65.

\bibitem{deledalle2009iwm}
C.~Deledalle, L.~Denis, and F.~Tupin, ``{Iterative Weighted Maximum Likelihood
  Denoising with Probabilistic Patch-Based Weights},'' \emph{IEEE Transactions
  on Image Processing}, vol.~18, no.~12, pp. 2661--2672, 2009.

\bibitem{parrilli_nonlocal_2012}
S.~Parrilli, M.~Poderico, C.~Angelino, and L.~Verdoliva, ``A {Nonlocal} {SAR}
  {Image} {Denoising} {Algorithm} {Based} on {LLMMSE} {Wavelet} {Shrinkage},''
  \emph{IEEE Transactions on Geoscience and Remote Sensing}, vol.~50, no.~2,
  pp. 606--616, Feb. 2012.

\bibitem{martino_scattering_based_nonlocal_2016}
G.~D. Martino, A.~D. Simone, A.~Iodice, and D.~Riccio, ``Scattering-based
  nonlocal means sar despeckling,'' \emph{IEEE Transactions on Geoscience and
  Remote Sensing}, vol.~54, no.~6, pp. 3574--3588, 06 2016.

\bibitem{deledalle_nl-insar:_2011}
C.-A. Deledalle, L.~Denis, and F.~Tupin, ``{NL}-{InSAR}: Nonlocal interferogram
  estimation,'' \emph{IEEE Transactions on Geoscience and Remote Sensing},
  vol.~49, no.~4, pp. 1441--1452, Apr. 2011.

\bibitem{lin_insar_tensor_svd_2015}
X.~Lin, F.~Li, D.~Meng, D.~Hu, and C.~Ding, ``Nonlocal sar interferometric
  phase filtering through higher order singular value decomposition,''
  \emph{IEEE Geoscience and Remote Sensing Letters}, vol.~12, no.~4, pp.
  806--810, 4 2015.

\bibitem{chen_nonlocal_polsar_pretest_2011}
J.~Chen, Y.~Chen, W.~An, Y.~Cui, and J.~Yang, ``Nonlocal filtering for
  polarimetric sar data: A pretest approach,'' \emph{IEEE Transactions on
  Geoscience and Remote Sensing}, vol.~49, no.~5, pp. 1744--1754, 5 2011.

\bibitem{deledalle_nl-sar:_2015}
C.-A. Deledalle, L.~Denis, F.~Tupin, A.~Reigber, and M.~Jäger, ``{NL}-{SAR}:
  {A} {Unified} {Nonlocal} {Framework} for {Resolution}-{Preserving}
  ({Pol})({In}){SAR} {Denoising},'' \emph{IEEE Transactions on Geoscience and
  Remote Sensing}, vol.~53, no.~4, pp. 2021--2038, Apr. 2015.

\bibitem{sica_nonlocal_multipass_2015}
F.~Sica, D.~Reale, G.~Poggi, L.~Verdoliva, and G.~Fornaro, ``Nonlocal adaptive
  multilooking in sar multipass differential interferometry,'' \emph{IEEE
  Journal of Selected Topics in Applied Earth Observations and Remote Sensing},
  vol.~8, no.~4, pp. 1727--1742, 04 2015.

\bibitem{hondt_nonlocal_tomosar_2018}
O.~D’Hondt, C.~López-Martínez, S.~Guillaso, and O.~Hellwich, ``Nonlocal
  filtering applied to 3-d reconstruction of tomographic sar data,'' \emph{IEEE
  Transactions on Geoscience and Remote Sensing}, vol.~56, no.~1, pp. 272--285,
  01 2018.

\bibitem{breit_itp_2010}
H.~Breit, T.~Fritz, U.~Balss, A.~Niedermeier, M.~Eineder, N.~Yague-Martinez,
  and C.~Rossi, ``Processing of bistatic {TanDEM-X} data,'' in \emph{Geoscience
  and Remote Sensing Symposium (IGARSS), 2010 IEEE International}, 7 2010, pp.
  2640--2643.

\bibitem{fritz_itp_2011}
T.~Fritz, C.~Rossi, N.~Yague-Martinez, F.~Rodriguez-Gonzalez, M.~Lachaise, and
  H.~Breit, ``Interferometric processing of {TanDEM-X} data,'' in
  \emph{Geoscience and Remote Sensing Symposium (IGARSS), 2011 IEEE
  International}, 7 2011, pp. 2428--2431.

\bibitem{rossi_tandemx_rawdem_2012}
\BIBentryALTinterwordspacing
C.~Rossi, F.~R. Gonzalez, T.~Fritz, N.~Yague-Martinez, and M.~Eineder,
  ``{TanDEM-X} calibrated raw {DEM} generation,'' \emph{ISPRS Journal of
  Photogrammetry and Remote Sensing}, vol.~73, pp. 12--20, 2012. [Online].
  Available:
  \url{http://www.sciencedirect.com/science/article/pii/S0924271612001062}
\BIBentrySTDinterwordspacing

\bibitem{krieger_tandemx_2007}
G.~Krieger, A.~Moreira, H.~Fiedler, I.~Hajnsek, M.~Werner, M.~Younis, and
  M.~Zink, ``Tandem-x: A satellite formation for high-resolution sar
  interferometry,'' \emph{IEEE Transactions on Geoscience and Remote Sensing},
  vol.~45, no.~11, pp. 3317--3341, 11 2007.

\bibitem{lachaise_eusar_tandemx_hdem_insar_update_2016}
M.~Lachaise and T.~Fritz.

\bibitem{zhu_improving_2014}
``Improving {TanDEM}-{X} {DEMs} by {Non}-local {InSAR} {Filtering},'' in
  \emph{{EUSAR} 2014; {Proc.} of}.

\bibitem{zhu_nldem_2017}
X.~X. Zhu, M.~Lachaise, G.~Baier, Y.~Shi, F.~Adam, and R.~Bamler, ``Potential
  and limits of non-local {InSAR} filtering for {TanDEM-X} high-resolution
  {DEM} generation.''

\bibitem{hoffmann_tdx_specs_2016}
J.~Hoffmann, M.~Huber, U.~Marschalk, A.~Wendleder, B.~Wessel, M.~Bachmann,
  B.~Bräutigam, T.~Busche, J.~H. González, G.~Krieger, P.~Rizzoli,
  M.~Eineder, and T.~Fritz, ``{TanDEM-X Ground Segment DEM Products
  Specification Document},'' German Aerospace Center, Tech. Rep., 2016.

\bibitem{baier_igarss_gpu_nonlocal_2016}
G.~Baier and X.~X. Zhu, ``{GPU}-based nonlocal filtering for large scale {SAR}
  processing,'' in \emph{2016 IEEE International Geoscience and Remote Sensing
  Symposium (IGARSS)}, 7 2016, pp. 7608--7611.

\bibitem{lebrun_denoising_cuisine_2012}
M.~Lebrun, M.~Colom, A.~Buades, and J.~M. Morel, ``Secrets of image denoising
  cuisine,'' \emph{Acta Numerica}, vol.~21, p. 475–576, 2012.

\bibitem{dabov_bm3d_2007}
K.~Dabov, A.~Foi, V.~Katkovnik, and K.~Egiazarian, ``Image denoising by sparse
  3-d transform-domain collaborative filtering,'' \emph{IEEE Transactions on
  Image Processing}, vol.~16, no.~8, pp. 2080--2095, 8 2007.

\bibitem{salmon_two-stage_denoising_2012}
J.~Salmon, R.~Willett, and E.~Arias-Castro, ``A two-stage denoising filter: The
  preprocessed yaroslavsky filter,'' in \emph{2012 IEEE Statistical Signal
  Processing Workshop (SSP)}, 8 2012, pp. 464--467.

\bibitem{barash_framework_nonlinear_diffusion_bilateral_2004}
\BIBentryALTinterwordspacing
D.~Barash and D.~Comaniciu, ``A common framework for nonlinear diffusion,
  adaptive smoothing, bilateral filtering and mean shift,'' \emph{Image and
  Vision Computing}, vol.~22, no.~1, pp. 73 -- 81, 2004. [Online]. Available:
  \url{http://www.sciencedirect.com/science/article/pii/S0262885603001732}
\BIBentrySTDinterwordspacing

\bibitem{weickert_review_nonlinear_diffusion_filtering_1997}
J.~Weickert, \emph{A review of nonlinear diffusion filtering}.\hskip 1em plus
  0.5em minus 0.4em\relax Berlin, Heidelberg: Springer Berlin Heidelberg, 1997,
  pp. 1--28.

\bibitem{winnemoeller_video_abstraction_2006}
H.~Winnemöller, S.~C. Olsen, and B.~Gooch, ``Real-time video abstraction,''
  \emph{ACM Transactions on Graphics (Proc. of SIGGRAPH'06)}, vol.~25, no.~3,
  pp. 1221--1226, 7 2006.

\bibitem{duval_bias-variance_nl_means_2011}
\BIBentryALTinterwordspacing
V.~Duval, J.-F. Aujol, and Y.~Gousseau, ``{A bias-variance approach for the
  Non-Local Means},'' \emph{{SIAM Journal on Imaging Sciences}}, vol.~4, no.~2,
  pp. 760--788, 01 2011. [Online]. Available:
  \url{https://hal.archives-ouvertes.fr/hal-00947885}
\BIBentrySTDinterwordspacing

\bibitem{lee_insar_additive_1998}
J.~S. Lee, K.~P. Papathanassiou, T.~L. Ainsworth, M.~R. Grunes, and A.~Reigber,
  ``A new technique for noise filtering of {SAR} interferometric phase
  images,'' \emph{IEEE Transactions on Geoscience and Remote Sensing}, vol.~36,
  no.~5, pp. 1456--1465, 9 1998.

\bibitem{bamler_insar_1998}
R.~Bamler and P.~Hartl, ``Synthetic aperture radar interferometry,''
  \emph{Inverse Problems}, vol.~14, pp. R1--54, 1998.

\bibitem{guarnieri_quick_dirty_coherence_estimator_1997}
A.~M. Guarnieri and C.~Prati, ``Sar interferometry: a ``quick and dirty''
  coherence estimator for data browsing,'' \emph{IEEE Transactions on
  Geoscience and Remote Sensing}, vol.~35, no.~3, pp. 660--669, 5 1997.

\bibitem{kervrann_adaptive_search_window_2008}
\BIBentryALTinterwordspacing
C.~Kervrann and J.~Boulanger, ``Local adaptivity to variable smoothness for
  exemplar-based image regularization and representation,'' \emph{International
  Journal of Computer Vision}, vol.~79, no.~1, pp. 45--69, 8 2008. [Online].
  Available: \url{https://doi.org/10.1007/s11263-007-0096-2}
\BIBentrySTDinterwordspacing

\bibitem{suo_local_fringe_2010}
Z.~Suo, Z.~Li, and Z.~Bao, ``A new strategy to estimate local fringe
  frequencies for insar phase noise reduction,'' \emph{IEEE Geoscience and
  Remote Sensing Letters}, vol.~7, no.~4, pp. 771--775, 10 2010.

\bibitem{fedorov_affine_nonlocal_2017}
V.~Fedorov and C.~Ballester, ``Affine non-local means image denoising,''
  \emph{IEEE Transactions on Image Processing}, vol.~26, no.~5, pp. 2137--2148,
  5 2017.

\bibitem{goodman_multivar_complex_gaussian_1963}
\BIBentryALTinterwordspacing
N.~R. Goodman, ``Statistical analysis based on a certain multivariate complex
  gaussian distribution (an introduction),'' \emph{Ann. Math. Statist.},
  vol.~34, no.~1, pp. 152--177, 03 1963. [Online]. Available:
  \url{https://doi.org/10.1214/aoms/1177704250}
\BIBentrySTDinterwordspacing

\bibitem{fournier_diamond_square_1982}
\BIBentryALTinterwordspacing
A.~Fournier, D.~Fussell, and L.~Carpenter, ``Computer rendering of stochastic
  models,'' \emph{Commun. ACM}, vol.~25, no.~6, pp. 371--384, Jun. 1982.
  [Online]. Available: \url{http://doi.acm.org/10.1145/358523.358553}
\BIBentrySTDinterwordspacing

\bibitem{rossi_paddy_rice_2015}
C.~Rossi and E.~Erten, ``Paddy-rice monitoring using tandem-x,'' \emph{IEEE
  Transactions on Geoscience and Remote Sensing}, vol.~53, no.~2, pp. 900--910,
  2 2015.

\bibitem{nolan_penetration_depth_2003}
M.~Nolan and D.~R. Fatland, ``Penetration depth as a dinsar observable and
  proxy for soil moisture,'' \emph{IEEE Transactions on Geoscience and Remote
  Sensing}, vol.~41, no.~3, pp. 532--537, 3 2003.

\bibitem{lee_speckle_analysis_1981}
\BIBentryALTinterwordspacing
J.-S. Lee, ``Speckle analysis and smoothing of synthetic aperture radar
  images,'' \emph{Computer Graphics and Image Processing}, vol.~17, no.~1, pp.
  24 -- 32, 1981. [Online]. Available:
  \url{http://www.sciencedirect.com/science/article/pii/S0146664X81800056}
\BIBentrySTDinterwordspacing

\end{thebibliography}

\begin{IEEEbiography}[{\includegraphics[width=1in,height=1.5in,clip,keepaspectratio]{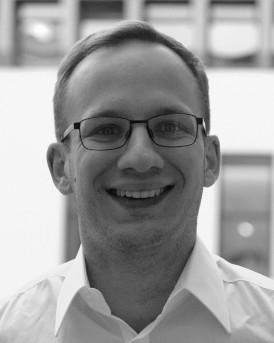}}]{Gerald Baier}
received the B.Sc. degree from the Karlsruhe Institute of Technology (KIT) in 2010, and the M.Sc. degree from the Université catholique de Louvain and the Karlsruhe Institute of Technology in 2012, both in electrical engineering.
In 2014 he joined the Remote Sensing Technology Institute of the German Aerospace Center (DLR) to pursue the Ph.D. degree in the field of synthetic aperture radar interferometry.
His research interests include signal processing, image denoising and high-performance computing.
\end{IEEEbiography}

\begin{IEEEbiography}[{\includegraphics[width=1in,height=1.5in,clip,keepaspectratio]{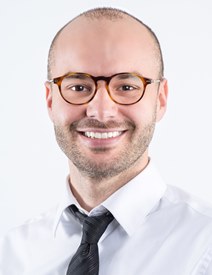}}]{Cristian Rossi}
received the B.Sc. and M.Sc. degrees in telecommunication engineering from the Polytechnic of Milan, Italy, and the Ph.D. in remote sensing technology from the Technical University of Munich, Germany, in 2003, 2006 and 2016, respectively.
From 2006 to 2008, he was Project Engineer with Aresys, Milan, Italy. From 2008 to 2017, he was Research Scientist with the German Aerospace Center (DLR), Oberpfaffenhofen, Germany, where he worked on the development of the integrated TanDEM-X processor and on novel interferometry algorithms for synthetic aperture radar missions. Since 2017, he is the Principal Earth Observation Specialist at the Satellite Applications Catapult, Harwell Campus, UK, where he is technically leading several national and international projects focused on the exploitation of remote sensing data for land applications. His research interests include urban remote sensing, multisource data fusion, digital elevation models, and environmental parameter estimation.

\end{IEEEbiography}

\begin{IEEEbiography}[{\includegraphics[width=1in,height=1.5in,clip,keepaspectratio]{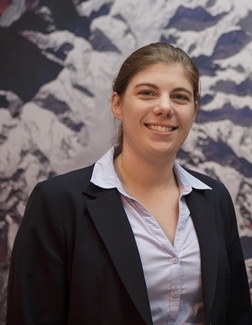}}]{Marie Lachaise}
received the Diploma degree in electronics, telecommunications and computer science from the Ecole Supérieure de Chimie, Physique, Electronique de Lyon, Villeurbanne, France, the M.Sc. degree in embedded systems and medical images from the Université Claude Bernard Lyon 1, Villeurbanne, in 2005, and a PhD in Engineering from the Technical University of Munich in 2015. 
Since 2005, she has been with the SAR Signal Processing Department, Remote Sensing Technology Institute, German Aerospace Center (DLR), Wessling, Germany. She developed software and algorithms for the TerraSAR-X and the TanDEM-X missions. Especially she designed the algorithms related to the interferometric processing of the multi-channel data of the TanDEM-X mission such as the multi-baseline phase unwrapping. Since 2017, she is SAR system engineers for the Tandem-L mission.

\end{IEEEbiography}

\begin{IEEEbiography}[{\includegraphics[width=1in,height=1.5in,clip,keepaspectratio]{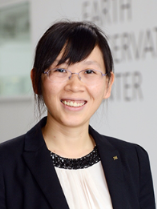}}]{Xiao Xiang Zhu}
(S'10--M'12--SM'14) received the Master (M.Sc.) degree, her doctor of engineering (Dr.-Ing.) degree and her “Habilitation” in the field of signal processing from Technical University of Munich (TUM), Munich, Germany, in 2008, 2011 and 2013, respectively.

She is currently the Professor for Signal Processing in Earth Observation (www.sipeo.bgu.tum.de) at Technical University of Munich (TUM) and German Aerospace Center (DLR); the head of the department of EO Data Science at DLR; and the head of the Helmholtz Young Investigator Group ”SiPEO” at DLR and TUM. Prof. Zhu was a guest scientist or visiting professor at the Italian National Research Council (CNR-IREA), Naples, Italy, Fudan University, Shanghai, China, the University  of Tokyo, Tokyo, Japan and University of California, Los Angeles, United States in 2009, 2014, 2015 and 2016, respectively. Her main research interests are
remote sensing and earth observation, signal processing, machine learning and data science, with a special application focus on global urban mapping.

Dr. Zhu is a member of young academy (Junge Akademie/Junges Kolleg) at the Berlin-Brandenburg Academy of Sciences and Humanities and the German National  Academy of Sciences Leopoldina and the Bavarian Academy of Sciences and Humanities. She is an associate Editor of IEEE Transactions on Geoscience and Remote Sensing.
\end{IEEEbiography}

\begin{IEEEbiography}[{\includegraphics[width=1in,height=1.5in,clip,keepaspectratio]{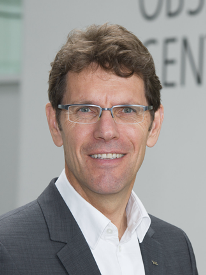}}]{Richard Bamler}
(M'95--SM'00--F'05) received his Diploma degree in Electrical Engineering, his Doctorate in Engineering, and his “Habilitation” in the field of signal and systems theory in 1980, 1986, and 1988, respectively, from the Technical University of Munich, Germany.

He worked at the university from 1981 to 1989 on optical signal processing, holography, wave propagation, and tomography. He joined the German Aerospace Center (DLR), Oberpfaffenhofen, in 1989, where he is currently the Director of the Remote Sensing Technology Institute.

In early 1994, Richard Bamler was a visiting scientist at Jet Propulsion Laboratory (JPL) in preparation of the SIC-C/X-SAR missions, and in 1996 he was guest professor at the University of Innsbruck. Since 2003 he has held a full professorship in remote sensing technology at the Technical University of Munich as a double appointment with his DLR position. His teaching activities include university lectures and courses on signal processing, estimation theory, and SAR. Since he joined DLR Richard Bamler, his team, and his institute have been working on SAR and optical remote sensing, image analysis and understanding, stereo reconstruction, computer vision, ocean color, passive and active atmospheric sounding, and laboratory spectrometry. They were and are responsible for the development of the operational processors for SIR-C/X-SAR, SRTM, TerraSAR-X, TanDEM-X, Tandem-L, ERS-2/GOME, ENVISAT/SCIAMACHY, MetOp/GOME-2, Sentinel-5P, Sentinel-4, DESIS, EnMAP, etc.

Richard Bamler’s research interests are in algorithms for optimum information extraction from remote sensing data with emphasis on SAR. This involves new estimation algorithms, like sparse reconstruction, compressive sensing and deep learning.
\end{IEEEbiography}

\end{document}